\documentclass[pdflatex,sn-basic]{sn-jnl}

\newcommand{\Deleted}[1]{}
\newcommand{\Edit}[1]{#1}

\usepackage{graphicx}%
\usepackage{multirow}%
\usepackage{amsmath,amssymb,amsfonts}%
\usepackage{amsthm}%
\usepackage{mathrsfs}%
\usepackage[title]{appendix}%
\usepackage{xcolor}%
\usepackage{textcomp}%
\usepackage{manyfoot}%
\usepackage{booktabs}%
\usepackage{algorithm}%
\usepackage{algorithmicx}%
\usepackage{algpseudocode}%
\usepackage{listings}%
\usepackage{eurosym}
\usepackage{acronym}

\usepackage{xcolor}

\theoremstyle{thmstyleone}%
\theoremstyle{thmstyletwo}%

\theoremstyle{thmstylethree}%

\acrodef{VR}[VR]{\emph{Virtual Reality}}
\acrodef{AR}[AR]{\emph{Augmented Reality}}
\acrodef{XR}[XR]{\emph{Extended Reality}}
\acrodef{EDA}[EDA]{\emph{Electrodermal Activity}}
\acrodef{HR}[HR]{\emph{Heart Rate}}
\acrodef{HMD}[HMD]{\emph{Head-Mounted Display}}
\acrodef{HRV}[HRV]{\emph{Heart Rate Variability}}
\acrodef{HPA}[HPA]{\emph{Hypothalamic-Pituitary-Adrenal}}
\acrodef{SSQ}[SSQ]{\emph{Simulator Sickness Questionnaire}}
\acrodef{FMS}[FMS]{\emph{Fast Motion Sickness Questionnaire}}
\acrodef{EEG}[EEG]{\emph{Electroencephalography}}
\acrodef{ECG}[ECG]{\emph{Electrocardiography}}
\acrodef{EMG}[EMG]{\emph{Electromyography}}
\acrodef{EOG}[EOG]{\emph{Electrooculography}}
\acrodef{EGG}[EGG]{\emph{Electrogastrography}}
\acrodef{VIMSSQ}[VIMSSQ]{\emph{Visually Induced Motion Sickness Susceptibility Questionnaire}}
\acrodef{RT}[RT]{\emph{Reaction Time}}
\acrodef{CV}[CV]{\emph{Coefficient of Variation}}

\begin{document}

\title[]{Do Not Immerse and Drive? Prolonged Effects of Cybersickness on Physiological Stress Markers And Cognitive Performance}


\author*[1]{\fnm{Daniel} \sur{Zielasko}}\email{daniel.zielasko@rwth-aachen.de}

\author[2]{\fnm{Ben} \sur{Rehling}}

\author[2,3]{\fnm{Bernadette} \sur{von Dawans}}\email{vondawans@uni-trier.de}

\author*[2,3]{\fnm{Gregor} \sur{Domes}}\email{domes@uni-trier.de}

\affil[1]{\orgdiv{Engineering Technology}, \orgname{Technical University of Denmark}, \country{Denmark}}

\affil[2]{\orgdiv{Department of Biological and Clinical Psychology}, \orgname{Trier University}, \country{Germany}}

\affil[3]{\orgdiv{Institute for Cognitive and Affective Neuroscience}, \orgname{Trier University}, \country{Germany}}

\abstract{Extended exposure to virtual reality environments can induce motion sickness, often referred to as cybersickness, which may lead to physiological stress responses and impaired cognitive performance. This study investigates the aftereffects of VR-induced motion sickness with a focus on physiological stress markers and working memory performance. Using a carousel simulation to elicit cybersickness, we assessed subjective discomfort (SSQ, FMS), physiological stress (salivary cortisol, alpha-amylase, electrodermal activity, heart rate), and cognitive performance (n-Back task) over a 90-minute post-exposure period. Our findings demonstrate a significant increase in both subjective and physiological stress indicators following VR exposure, accompanied by a decline in working memory performance. Notably, delayed symptom progression was observed in a substantial proportion of participants, with some reporting peak symptoms up to 90 minutes post-stimulation. Salivary cortisol levels remained elevated throughout the observation period, indicating prolonged stress recovery. These results highlight the need for longer washout phases in XR research and raise safety concerns for professional applications involving post-exposure task performance.}





\maketitle

\section{Introduction}
\ac{VR} and \ac{AR} technologies are increasingly being integrated into cognitively demanding work environments, such as aviation, surgery, and complex manufacturing processes. 
These \ac{XR} technologies offer immersive training opportunities, real-time data visualization, and enhanced decision-making support. 
However, despite its advantages, prolonged exposure to \ac{XR} can introduce physiological and cognitive challenges, including motion sickness, fatigue, and stress responses, which may impact performance and well-being \citep{Parsons2018VRChallenges, Rizzo2017clinical}. 
Understanding how these environments affect users is crucial for optimizing their application in high-stakes professional settings.

A key issue in immersive environments is cybersickness, a form of motion sickness induced by sensory discrepancies between visual and vestibular signals \citep{Reason1975motion}. 
It manifests as nausea, dizziness, headache, and other autonomic symptoms frequently experienced by VR users \citep{Dennison2016, Keshavarz2015}.
Motion sickness in VR can occur under various conditions, including prolonged exposure, high visual flow, latency between head movements and visual updates, and discrepancies between acceleration cues and expected motion \citep{Stanney1997}. 
The severity and duration of symptoms vary among individuals, with some experiencing mild discomfort that subsides within minutes, while others report persistent symptoms lasting hours after VR exposure \citep{LaViola2000}. 
Factors such as individual susceptibility, prior \ac{VR} experience, and display technology play significant roles in determining the intensity of motion sickness effects.

The washout phase of sickness after \ac{XR} exposure has been studied, though not in great depth. 
However, it remains a critical aspect to explore further, particularly given its implications for both the robustness of scientific methods \citep{zielasko2024carryOver} and safety in high-stakes environments. 
For instance, understanding how long symptoms persist and how quickly they subside is essential when considering the use of \ac{XR} technologies in professions where precise performance is required, such as surgery or aviation. 
Prolonged exposure to immersive technologies may lead to lingering symptoms that could impair a user’s ability to operate potentially dangerous machinery or perform high-risk tasks. 
As \ac{XR} technologies become more prevalent in work environments, studying recovery from sickness is vital to ensure safe and effective usage in critical settings.

Previous studies have demonstrated that cybersickness can elicit a stress response, reflected in autonomic and endocrine changes. 
Research has shown that individuals experiencing higher discomfort in VR environments exhibit increased cortisol levels, altered \ac{EDA}, and \ac{HR} patterns \citep{Bos2005motion, Munafo2017TheVR}. 
A recent study by \cite{Kim2022} further supports this by demonstrating that cybersickness caused by \ac{HMD}-based \ac{VR} environments significantly influences physiological responses, including elevated \ac{HRV}, altered autonomic nervous system activity, and increased cortisol levels. 
The physiological stress response is characterized by the activation of the \ac{HPA} axis and the sympathetic nervous system \citep{McEwen2007}. 
While the \ac{HPA} axis regulates cortisol release, the sympathetic nervous system governs rapid autonomic adaptations, manifesting in increased \ac{EDA} and changes in \ac{HR} \citep{ulrich2009neural}.

Stress has been shown to impact cognitive functions, particularly working memory. 
While acute stress can enhance cognitive performance under specific conditions \citep{Sandi2013}, extensive research suggests that high stress levels impair working memory performance \citep{Schoofs2008, Shields2016}. 
This impairment may be due to increased cortisol and norepinephrine secretion, which modulate neural networks in the prefrontal cortex, thereby affecting working memory function \citep{Arnsten2009StressPrefrontal}.

Based on these findings, the present study examines the effects of motion sickness in a VR environment on physiological and cognitive parameters. 
Specifically, we investigate how \ac{VR}-induced motion sickness progresses after stimulation and whether it elicits a physiological stress response, and whether this response affects working memory performance. 
We hypothesize that \ac{VR} stimulation (carousel ride) leads to a heightened physiological stress response compared to the resting condition, as measured by cortisol and alpha-amylase concentrations, \ac{EDA}, and \ac{HR}. 
Furthermore, we predict that cognitive performance in the working memory task (\emph{n-Back} task) is significantly reduced following \ac{VR} stimulation compared to the resting condition. 

This study extends previous research by incorporating multiple physiological markers, allowing for a more comprehensive understanding of the motion sickness stress response in VR environments. 
By examining these effects in a controlled experimental setting, this research contributes to a deeper understanding of the interaction between VR, motion sickness, stress, and cognitive performance.
\Edit{
Given the limited and heterogeneous evidence on prolonged post-exposure effects of cybersickness across multiple physiological systems, the present study adopts an exploratory approach to characterize the temporal dynamics of subjective, physiological, and cognitive responses.
}

\section{Related Work\label{sec_rw}}
\subsection{Temporal Progression of Cybersickness During Exposure}
Existing research offers diverse perspectives on the temporal dynamics of cybersickness symptoms during exposure to virtual environments.

Several studies indicate that cybersickness tends to increase progressively with prolonged exposure.
\cite{Kennedy2000duration} observed a linear rise in symptoms among fighter pilots exposed to VR conditions over three hours. 
Similarly, \cite{Serge2015simulator} reported an upward trend in symptoms over a 16-minute duration, while \cite{Risi2019effects} found a comparable linear increase within the first ten minutes. 
\cite{McHugh2019} noted a similar trajectory in symptoms over approximately six minutes. \cite{zielasko2018dynamic}, who conducted self-reported assessments across two 15-minute conditions, also found indications of a continuous increase.

Longer-duration studies reinforce this trend. 
\cite{Stanney2003} examined exposure durations of 15, 30, 45, and 60 minutes in a between-subjects design, identifying a sustained increase in cybersickness beyond the 15-minute mark. 
\cite{Lampton1994} further corroborated this finding, showing that longer exposures correlated with more intense symptoms.

A meta-analysis by \cite{Saredakis2020factors} categorized studies into exposure groups of 0-10 minutes, 10-20 minutes, and beyond 20 minutes. 
Their findings indicated stable oculomotor symptoms across these groups, whereas nausea and disorientation peaked between 10 and 20 minutes. 
However, this contrasts with other research suggesting a sustained increase in symptoms beyond 20 minutes.

Some studies suggest adaptation effects over prolonged exposure. 
In a two-hour experiment, symptoms plateaued after 75 minutes, hinting at potential habituation \citep{Cobb1999}. 
\cite{Melo2018} found no significant \ac{SSQ} \citep{Kennedy1993} score differences between early and later exposure stages, reinforcing the notion of early adaptation. 
Similarly, studies using the \ac{FMS} \citep{Keshavarz2011validating} values \citep{McHugh2019, Thorp2022} primarily within the first ten minutes reported adaptation trends.

\cite{Sinitski2018} investigated cybersickness in treadmill-based virtual experiences, noting a significant increase in \ac{SSQ} disorientation scores up to 15 minutes, followed by an adaptation phase at 45 minutes. 
Interestingly, nausea peaked at 45 minutes, aligning with \cite{Hakkinen2018time}'s findings, whereas oculomotor symptoms plateaued after 15 minutes. 
Their data suggest a more complex progression where adaptation occurs alongside symptom escalation.

Other research challenges the notion of a purely gradual increase in symptoms, instead proposing an episodic pattern \citep{Wang2019, Zielasko2021sickness, Teixeira2022unexpected}. 
Such fluctuations may be difficult to detect due to limitations in sampling rates, which are often constrained by practical feasibility.

Taken together, these findings highlight both a general trend of increasing cybersickness over time and potential adaptation effects at various stages of exposure, emphasizing the need for finer-grained temporal analyses to capture the nuanced progression of symptoms.

\subsection{Temporal Progression of Cybersickness Post Stimulation\label{sec_rw_postexposure}}
Research is sparse on how cybersickness reacts post-stimulation and shows significant interindividual differences in this aspect. 
For instance, while most subjects experience minimal symptoms after a 10-60 minute break, some continue to suffer from severe symptoms~\citep{Singer1998}.

A key issue in cybersickness research is the lack of uniformity in post-exposure studies. 
The need for standard methodologies in this area has been discussed, with suggestions for establishing specific standards \citep{Duzmanska2018}. 

In a study measuring \ac{SSQ} 30 minutes after \ac{VR} exposure, no significant differences were found compared to a pre-study baseline, suggesting rapid recovery in most subjects. 
However, the disorientation subscale remained significantly higher than pre-experiment levels \citep{Singer1998}.
Contrastingly, \cite{Gavgani2017}'s study indicates more prolonged effects of cybersickness. 
They measured cybersickness before, immediately after, and one, two, and three hours post-exposure over three days. 
\ac{SSQ} values were significantly higher post-exposure each day, challenging the notion of rapid recovery suggested by Singer et al. 
Yet, both studies found significantly lower \ac{SSQ} values in the first follow-up compared to immediately post-exposure.
Further, recovery patterns vary with the VR experiment's format. 
In experiments with high \ac{SSQ} scores, some participants needed over 30 minutes to recover. 
In contrast, others reported no discomfort just five minutes post-experiment \citep{Tanaka2004virtual}.
This observation was mirrored by \cite{Szpak2020exergaming}, where most of the participants no longer felt any symptoms 40 minutes after exposure, whereas 14\% still felt strong after-effects.

Few studies have examined participants' well-being the day after the experiment. 
For example, a study with pilots post-simulator training revealed that 4.6\% reported symptoms 24 hours after training \citep{Ungs1987}. 
Another study with pilots showed that those with intense symptoms had significantly higher values two hours post-experiment compared to a less affected group, but no differences were observed after six, 12-18, and 24 hours \citep{Stoffregen2000}.

In summary, cybersickness symptoms typically last 1-2 hours but can persist for days in extreme cases \citep{Ilyas2012, Rebenitsch2016, Tyrrell2018, Zielasko2021sickness}.
Additionally, numerous studies consistently demonstrate an association between habituation or adaptation to cybersickness and repeated exposure to \ac{VR} applications \citep{Ilyas2012, Rebenitsch2016, Tyrrell2018, Zielasko2021sickness}. 
There is much less knowledge about the precise models governing the decay over time. 

\subsection{Cybersickness and Biophysical Responses}
Traditionally, subjective measures such as the \ac{SSQ} and the \ac{FMS} have been used to assess the severity of cybersickness. 
However, these self-reported tools have limitations, including relying on users' subjective perception, which can be influenced by factors such as mood, prior VR experience, and individual sensitivity. 
Consequently, there has been growing interest in the use of biophysical measures as \Edit{complementary} objective indicators of cybersickness.

\Deleted{
Biophysical measures, such as \ac{EEG}, \ac{ECG}, \ac{EDA}, and \ac{HRV}, offer the potential to complement or even replace subjective measures.
}
\Edit{
Biophysical measures, such as \ac{EEG}, \ac{ECG}, \ac{EDA}, and \ac{HRV}, have therefore been explored as potential complementary sources of information alongside subjective measures.
}
These physiological signals provide insights into the autonomic nervous system’s response to \ac{VR} stimuli, \Edit{which may reflect physiological changes associated with cybersickness}. 
\Deleted{
making them valuable for more accurately assessing the onset and severity of cybersickness.
}
\Edit{
However, current evidence suggests that no single biophysical measure can yet serve as a reliable standalone indicator of cybersickness across contexts.
}
Additionally, biophysical signals may enable the development of real-time systems that predict and mitigate cybersickness, \Edit{although such approaches typically rely on multimodal data and context-specific models rather than individual signals alone \citep{Keshavarz2022}.}

Recent studies have demonstrated that these biophysical measures can capture physiological responses that are associated with cybersickness under specific experimental conditions. 
For instance, changes in \ac{HR}, skin conductance level, and gastric activity have been reported to correlate with cybersickness severity in certain setups \citep{Dennison2016, Garcia2019development}. 
\Deleted{
Similarly, EEG measures, particularly changes in brainwave patterns, have been identified as reliable indicators of cybersickness.
}
\Edit{
Similarly, several studies have reported associations between specific \ac{EEG}-derived features (e.g., spectral power in selected frequency bands) and cybersickness-related outcomes \citep{Li2020, Liao2020, Kim2019}.
}
\Deleted{
These findings suggest that biophysical measures could effectively supplement subjective measures like the SSQ, offering more accurate, continuous, and objective assessments of cyber sickness.
}
\Edit{
Overall, these findings suggest that biophysical measures may supplement subjective measures such as the SSQ by providing continuous, objective data streams, although their effectiveness appears to be highly dependent on the chosen features, indices, and experimental context.
Importantly, reported associations between biophysical measures and cybersickness typically refer to specific indices or features derived from physiological signals (e.g., spectral EEG power in selected frequency bands, heart rate metrics, or statistical properties of the tonic skin conductance level), rather than to the raw signals themselves. Accordingly, several studies emphasize the importance of specifying which indices are examined when interpreting such associations.
}

\cite{Recenti2021toward} employed a combination of \ac{EEG}, \ac{EMG} from the gastrocnemius muscles, and \ac{HR} to assess motion sickness, finding that specific \ac{EMG}-derived parameters were more informative than the selected \ac{EEG} features for predicting motion sickness. 
In a similar vein, \cite{Lin2018} demonstrated that increased power in the gamma band ($>$32Hz) of \ac{EEG}, coupled with elevated heart rate, was associated with heightened cybersickness severity.

The role of multiple physiological signals in predicting cybersickness has been explored in several studies. 
\cite{Dennison2016} utilized \ac{EMG}, \ac{EDA}, \ac{EOG}, \ac{ECG}, and respiration to study cybersickness and reported that specific indices of gastric activity, blink behavior, and breathing patterns were associated with cybersickness severity. 
Likewise, \cite{Garcia2019development} collected data on \ac{ECG}, \ac{EOG}, respiratory signals, and skin conductance from 66 participants and showed that selected features derived from these signals were informative for cybersickness assessment within their experimental setup.

\cite{Islam2020} compared \ac{HR} and \ac{EDA} signals between high and low cybersickness groups, finding significant group differences in heart rate, while no significant differences were observed for breathing patterns or the examined \ac{HRV} indices. 
\Deleted{
These results suggest that \ac{HR} is a more reliable indicator of cybersickness than other measures like respiration rate or \ac{HRV}.
}
\Edit{
These results indicate that heart rate-related measures may be more sensitive to cybersickness-related changes than the specific respiration and \ac{HRV} indices considered in that study.
}

\cite{Li2020} proposed a method for recognizing \ac{VR} motion sickness using \ac{EEG} rhythm energy ratios, achieving high recognition accuracy for individual subjects using a wavelet packet transform. 
Their method combines selected \ac{EEG} features with machine learning models such as support vector machines, highlighting a broader trend toward computational approaches for cybersickness detection.

\cite{Magaki2019} explored how specific \ac{HRV} and \ac{EDA} metrics during VR tasks could be used as potential indicators of cybersickness. 
Similarly, \cite{Dennison2019} employed a multi-modal approach incorporating \ac{EEG}, \ac{ECG}, \ac{EGG}, and postural sway, reporting that the selected \ac{EEG}-derived features yielded the highest predictive performance within their model.

More recently, \cite{Islam2022towads} introduced a multimodal deep fusion network combining \ac{HR}, \ac{EDA}, eye movements, and head-tracking data to predict the onset of cybersickness in real time. 
These advances reflect a broader trend in which research increasingly focuses on multimodal and machine learning-based approaches rather than treating individual biophysical signals as standalone indicators. 
Similarly, \cite{oh2021machine} proposed machine learning models using physiological and behavioral features, such as heart rate and body movements, achieving promising results for cybersickness onset prediction.

\cite{Kim2022} simulated aircraft motion in \ac{VR} and measured physiological responses including heart rate, blood pressure, body temperature, and cortisol levels. 
Their results showed increases in heart rate and cortisol levels associated with cybersickness, while blood pressure decreased. 
\Deleted{
The study concludes that \ac{HR} and cortisol are promising physiological markers for cybersickness assessment
}
\Edit{
The authors suggest that heart rate and cortisol may serve as useful physiological correlates of cybersickness within simulated flight scenarios
}
and call for future research on factors specifically affecting disorientation.

In conclusion, while there is substantial research examining associations between biophysical responses and cybersickness, most studies emphasize multimodal and machine learning-based prediction approaches. 
\Edit{
Consistent with recent reviews \citep{Keshavarz2022}, there is currently limited evidence that any individual biophysical measure or index can reliably and robustly indicate cybersickness across different VR contexts.
}
The integration of multiple biosignals shows promise, but the individual contributions of each signal remain underexplored, and further research is needed to better understand post-exposure physiological responses.

\section{Method\label{sec_method}}
To address our research questions, particularly regarding the progression of cybersickness symptoms after stimulation, we deliberately induced cybersickness in participants (study condition) and subsequently monitored them using physiological and endocrinological measures, established subjective scales such as the \ac{SSQ} and \ac{FMS}, and a cognitive load task. 
Participants were observed for up to 90 minutes post-stimulation.

To normalize individual differences and properly contextualize, we paired this design with a sequence-balanced within-subject control condition in which no sickness stimulus was applied. 
The study was reviewed and approved by the university’s ethics board.
It was conducted in line with the Declaration of Helsinki.

The following sections provide a detailed overview of the study methodology.
The study design was preregistered via \href{https://osf.io/7nxsj}{OSF}.

\subsection{Stimulus (IV) \& Apparatus\label{stimulus}}
\begin{figure}
    \centering
    \includegraphics[width = .8\columnwidth]{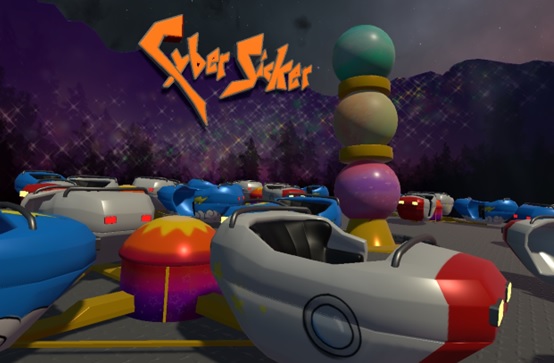}
    \caption{Egocentric view for the participants in the movement condition sitting in one of the Cybersicker carts.\label{fig_sickerEgo}}
\end{figure}

To induce a cybersickness stimulus, we utilized an open-access simulator developed by \cite{zielasko2024cybersicker}. 
This simulator immerses users in the perspective of a seated passenger in a cart on an amusement ride.

It is designed to provide a more controlled stimulus than typical virtual roller coasters (cf. \cite{Ang2023}), while still maintaining a realistic and fun experience, as opposed to highly abstract approaches such as an optoelectronic drum (cf. \cite{Bubka2006Optokinetic}).

Throughout the experience, participants remained physically seated. 
There was no interaction beyond the ability to look around freely. 
Their only task was to experience and observe the ride passively.
The simulator is modeled after a real amusement ride known as the \textit{Brakedancer} and is implemented in Unity. 
The base platform of the simulator is tilted by 8 degrees. 
We used no virtual body representation in the study.
Each cart is grouped with three others around a central pivot point. 
Individual carts can rotate around their axes, specifically the yaw and pitch axes. 
Additionally, all cart groups on the tilted platform collectively rotate around the platform’s central axis, which is aligned with a central tower structure.

At the start of the simulation, all three rotational axes begin moving at a slow pace. 
To ensure participant safety, the experiment was equipped with a dead man's switch, allowing the simulation to be halted at any time if a participant reported reaching 70\% of their cybersickness threshold (see Section~\ref{sec_fms}).

The amusement ride is set within a virtual fairground environment, which includes typical elements such as a free-fall tower, various booths, and a Ferris wheel.
During the simulation, both the participant's cart and the surrounding gondolas begin to rotate, gradually increasing in speed according to a predefined automatic temporal protocol (every 30 seconds). 
The complete parameterization is accessible with the publicly available source code.
The experience is presented from a first-person perspective in \ac{VR}. 
To deliver the stimulus, we employed an HTC VIVE Pro headset.

\subsection{Procedure}
\begin{figure}
    \centering
    \includegraphics[width = 1.\columnwidth]{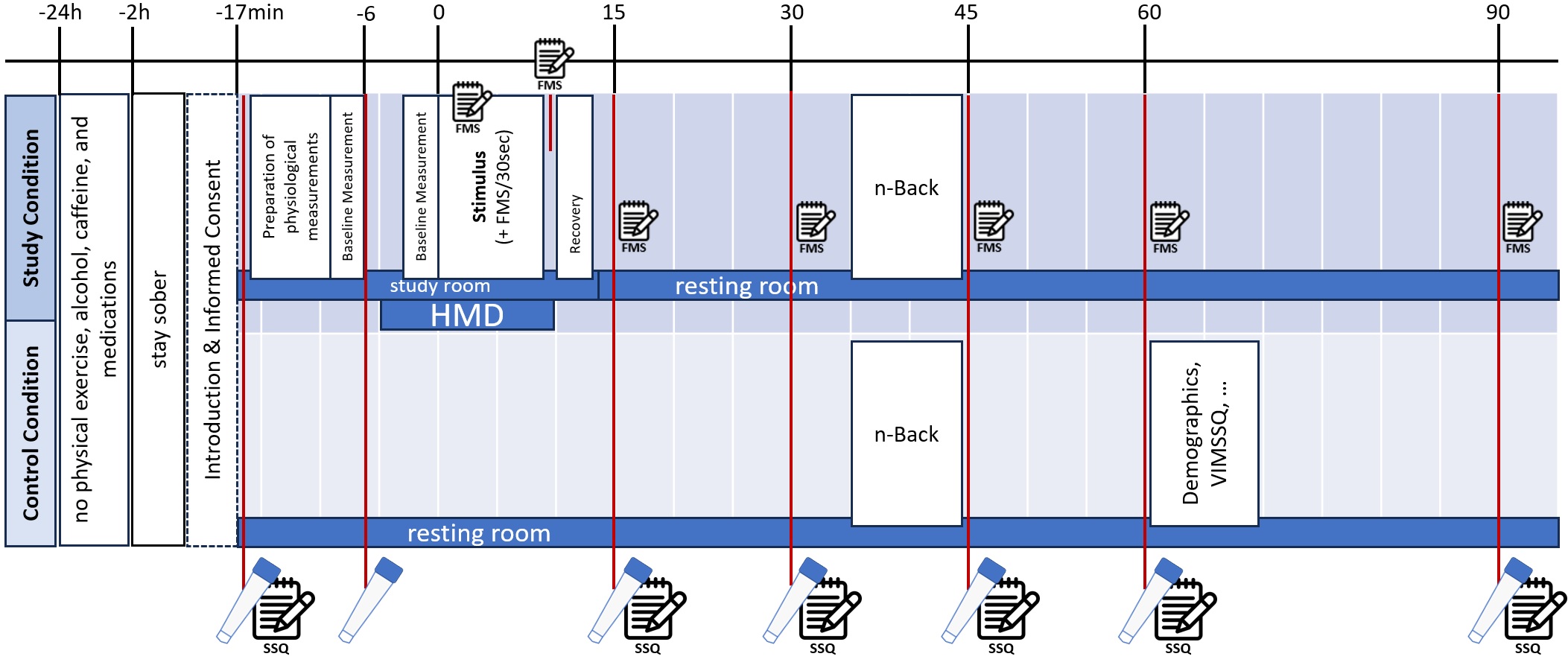}
    \caption{\Edit{Schematic representation of the study procedure. The experiment followed a sequence-balanced within-subject design with control and study sessions conducted on separate days.}\label{fig_procedure}}

\end{figure}

Participants were informed during recruitment that the experiment aimed to induce cybersickness. 
Each participant was invited to two afternoon sessions scheduled at the same time but several days apart. 
This timing was crucial for endocrinological measurements, as hormone levels exhibit significant fluctuations throughout the day.
For the same reason, participants were instructed to refrain from physical exercise, alcohol, caffeine, and medications at least 24 hours before testing, as well as to refrain from consuming anything but water two hours prior to testing. 
Additionally, participants agreed to hand in their mobile phones for the duration of the two-hour experiment to prevent uncontrolled influences on their stress levels, such as those caused by accessing news or messages.

Upon arrival at the first session, participants received both written and verbal explanations of the experiment. 
They then signed an informed consent form before being randomly assigned to a \Edit{sequence-balanced order of the two experimental sessions: a control condition without a cybersickness stimulus and a study condition in which the stimulus was presented (see Section~\ref{stimulus}).}
A schematic representation of the experimental procedure is shown in Fig.~\ref{fig_procedure}.

\subsubsection{Study Condition}
On the study condition day of the experiment, the participant’s interpupillary distance was measured using the GlassesOn app and adjusted in the \ac{HMD} to ensure an accurate \ac{VR} presentation. 
Subsequently, participants completed the \ac{SSQ} for the first time, assessing their current symptoms related to cybersickness. 
The experimenter then introduced the \ac{FMS} scale, explaining that a score of 0 represented no symptoms, while a score of 10 indicated severe sickness, such as nausea leading to vomiting. 
Participants were informed that during the sickness stimulus, they would be asked every 30 seconds to report their current sickness level via \ac{FMS}. 
If they reported a score of 7 or higher, the experiment would be stopped immediately, and they were made aware of this.
We did not aim to maximize sickness in this study, as participants needed to remain able to continue with the experiment and subsequent observations after the stimulus.

After introducing participants to the stimulus, the \ac{SSQ}, and the \ac{FMS}, the experimenter collected the first saliva sample ($t=-17$, where $t=0$ denotes the start of the stimulation). 
Following this, physiological measurements were prepared. 
Participants were fitted with a chest strap to monitor respiration, and ECG electrodes were placed in a triangular formation on both clavicles and the lower left side of the thoracic cage. 
The electrode sites were cleaned with alcohol (propanol) before placement. 
Electrodes for measuring \ac{EDA} were attached to the participant’s non-dominant hand and secured with tape. 
These physiological measurements were only recorded in the experimental condition and served as control variables.

To conclude the preparation phase, the first baseline psychophysiological activity was recorded for three minutes, followed by a second saliva baseline measurement ($t=-6$). 
Participants were then introduced to the VR environment, and after familiarization, another psychophysiological baseline was recorded for 3 minutes. 
Before starting the stimulation at $t=0$, the experimenter reiterated the termination criterion: a self-reported \ac{FMS} score of 7 or higher.

The \ac{VR} stimulation lasted either for a maximum of eight minutes or until the participant reached the termination threshold. 
After the stimulus ended either way, participants remained in the VR environment for an additional minute and were then asked to report their \ac{FMS} score. 
The HMD was then removed, followed by the final 3-minute psychophysiological measurement—the recovery phase—and another saliva sample ($t=15$).

After completing this phase, participants moved to a separate resting room. 
There, additional saliva samples, \ac{SSQ}, and \ac{FMS} assessments were collected at predefined time points ($t=30$, $45$, $60$, and $90$). 
At $t=35$, participants completed an \emph{n-Back} task. 
The resting room was designed to be a low-stimulation environment, providing a quiet space where participants could read pre-selected magazines or their own literature.

\subsubsection{Control Condition}
The control condition is particularly important for the endocrinological markers, as these are susceptible to environmental influences. 
Moreover, it became evident that subjective measurements such as the \ac{SSQ} do not fully adhere to the zero-baseline assumption \citep{Brown2022zeroBaseline}. 
As a result, these measurements also become more robust when an appropriate control condition is implemented.
\Deleted{
We opted for a low-level (passive) control condition rather than closely mimicking the experimental condition. 
This decision was made to control for the well-known circadian rhythm of the HPA axis. 
}
\Edit{
We deliberately chose a low-level (passive), time-matched control condition because our primary outcomes included endocrine markers related to HPA-axis activity, which are highly sensitive to contextual arousal, novelty, expectancy, and environmental stimulation. Any VR exposure — even a “less sickening” or stationary version — would still constitute an active manipulation (e.g., visual stimulation, immersion, novelty, equipment setup), potentially influencing HPA-axis responses independent of cybersickness. The passive control was therefore intended to provide a conservative baseline while controlling for the well-known circadian rhythm of the HPA axis \citep{Kirschbaum1994} as well as repeated sampling effects. Accordingly, participants in the control session followed the identical measurement schedule (saliva sampling and SSQ assessments at the same time points as in the study session) while resting in the recovery room.
}
\Deleted{
Participants in the control group were immediately directed to the resting room, where saliva samples and \ac{SSQ} assessments were taken at the same time points as in the experimental condition.
}
\Edit{
During the control session, participants were immediately directed to the resting room, where saliva samples and \ac{SSQ} assessments were taken at the same time points as in the study session.
} 
The first sample was labeled $t=-17$, following the notation of the study condition.
They also completed the \emph{n-Back} task at $t=35$.

\subsection{Measures (DV)\label{measures}}
\subsubsection{Questionnaires}
\paragraph{Simulator Sickness Questionnaire (SSQ)}
The participants filled out the \acf{SSQ} at six time points, once before the experiment ($t = -17$) and five times following the experiment ($t = 15$, $30$, $45$, $60$, and $90$). 
All timestamps refer to the cortisol measurement, while the \ac{SSQ} assessment always took place immediately before it, with a time offset of approximately 2 minutes.
This instrument measures the symptoms related to cybersickness of nausea, disorientation, and oculomotor problems on a 4-point Likert scale from None to Severe using 16 items. 
It provides the three subscales of nausea, disorientation, and oculomotor as well as an overall score of cybersickness.

\paragraph{Fast Motion Sickness Scale (FMS)\label{sec_fms}}
Further, the participants were asked to rate their feelings of comfort throughout the study condition on a scale from 0 (not sick at all) to 10 (frank sickness).
This scale is an adapted version of the \acf{FMS}. 
It was adapted from 0 to 20 to make it more intuitive for the participants under the potentially stressful circumstances this experiment evokes (cf. \citep{zielasko2016hmdnav, Adhikari2022}). 
The \ac{FMS} was used during the experiment to evaluate the current well-being of the participants during the stimulation. 
For this purpose, participants were asked to rate their current sickness every 30 seconds during the motion stimulation.
Once they reached the threshold of 7, the stimulation was terminated to prevent the participants from severe physical reactions such as throwing up. 
One minute after the experiment, as well as with every \ac{SSQ} following (study condition only), the participants gave another rating on the \ac{FMS}. 

\paragraph{Visually Induced Motion Sickness Susceptibility Questionnaire (VIMSSQ)}
Furthermore, the participants were asked to fill out the \acf{VIMSSQ} \citep{Keshavarz2023vimssq} in the control condition between $t = 60$ and $90$. 
By using this instrument, we expect to evaluate the susceptibility of the participants to show symptoms related to motion sickness while using visual displays. 
It is expected to predict the \ac{SSQ} values that the participants provide following the cybersickness-inducing experiment.  

\paragraph{Demographic Questionnaire}
We prepared a demographic questionnaire, asking the participants about their sex, age, weight, height, physical activities, their experience in 3D video games, their experience in VR environments, their subjective susceptibility to motion sickness, how they respond to stress as well as whether they expect to experience cybersickness.  

\subsubsection{Salivary Cortisol and Alpha-Amylase}
Saliva samples were collected at seven time points throughout the experiment ($t=-17$, $-6$, $15$, $30$, $45$, $60$, $90$). 
The samples were collected by the participants themselves under supervision and with instructions on the proper procedure. 
They were given a cotton swab, which they moved around the inside of their mouth for 45–60 seconds, focusing particularly on the cheeks. 
Afterward, they placed the swab into the designated and labeled tube and handed it over to the experimenter.
The samples were initially stored at the university facility where the experiment was conducted and later transported to the laboratory. 
At both locations, they were preserved at -18°C until biochemical analysis.

For this experiment, we used Salivettes (Sarstedt, Nümbrecht, Germany). 
Salivary cortisol levels were measured using a time-resolved fluorescence immunoassay \citep{Dressendorfer1992CortisolBiotin}. 
Additionally, alpha-amylase activity was determined using 2-chloro-4-nitrophenyl-D-maltotriose (CNP-G3) as the relevant substrate.

The enzymatic action of alpha-amylase on this substrate yields 2-chloro-p-nitrophenol, which can be spectrophotometrically measured at 405 nm. 
The amount of alpha-amylase activity present in the sample is directly proportional to the increase in absorbance at 405 nm \citep{LorentzGutschowRenner1999, WinnDeen1988}. 
The intra-assay \ac{CV} was 4.0-6.7\%, the corresponding inter-assay \ac{CV} was 7.1-9.0\% for cortisol 2.8 - 6.3\% (intra), and 5.5 – 7.6\% (inter) for sAA. 
Analyses were performed in duplicate and averaged.

\subsubsection{Psychophysiological Measures}
All psychophysiological measures were recorded between approximately $t=-5$ and $15$ minutes of the study condition with an integrated multichannel amplifier (Brainproducts V-Amp 16, Gilching, Germany) attached to a standard PC. 
Signals and section markers were recorded with Brainvision Recorder software (Brainproducts, Gilching, Germany) at a sampling rate of 1000 Hz and exported to EDF files for further preprocessing and analyses.

\paragraph{Heart Rate (HR) \& Heart Rate Variability (HRV)}
To measure \acf{HR} and calculate the \acf{HRV} throughout the experiment, a single-channel \ac{ECG} was recorded. 
Disposable electrodes were attached beneath the right collarbone and below the left ribcage (cf. Einthoven II). 
Peak detection and semiautomatic artifact rejection were done with Kubios HRV Scientific (Kubios, Kuopio, Finland). 
Kubios was also used for the segmentation and calculation of average \ac{HR} and \ac{HRV} RMSSD for the baseline, motion, and recovery periods.

\paragraph{Electrodermal Activity (EDA)}
The \acf{EDA}, more specific the \emph{Skin Conductance Level}, was recorded from the index and ring finger of the non-dominant hand using Ag/AgCl electrodes (4 mm diameter) filled with a conductive paste for \ac{EDA} recordings (Neurospec AG, Stans, Switzerland). 
This raw signal was low-pass filtered (0.3 Hz, Butterworth, 4th order), and segmented and averaged (median) to calculate SCL for the baseline, motion, and recovery period.

\subsubsection{n-Back Task}
Participants have completed a series of \emph{n-Back} tasks at $t = 35$ in both conditions.
The \emph{n-Back} task consisted of three blocks with ascending working memory load (1-back to 3-back). 
In each block, 60 items (14 different capital letters, 15 target stimuli) were presented. 
Each trial started with a blank screen (1000 ms) followed by a letter (max. 2000 ms until response or time out). 
For each stimulus presented, participants had the task of indicating by keystroke whether it had been presented one, two, or three times previously or not. 
The stimuli were presented centrally on a PC screen (specify), and the responses were recorded using a standard PC keyboard with two color-coded keys. 
Each block lasted about 3 minutes. 
The presentation was made using PsychoPy (insert current version and citation). Correct vs. incorrect responses and reaction times were recorded for subsequent statistical analysis.

\subsection{Participants\label{participants}}
\Deleted{
According to Kim et al., we expected a medium effect size of $.7$ in the psycho-endocrinological responses. 
To calculate the required sample size, we used G*power 3.1, which resulted in a minimum of $N = 24$ participants ($\alpha = .05$, $1-\beta = .95$).
}
\Edit{
Given the limited and heterogeneous evidence on psycho-endocrinological and psychophysiological correlates of cybersickness, this study was designed as largely exploratory. The sample size was therefore determined based on prior related work rather than a single confirmatory hypothesis. Following \cite{Kim2022}, we used G*Power 3.1 \citep{Faul2007gPower} to obtain an initial orientation for the required sample size, assuming a medium effect size ($d = .7$). This resulted in a minimum sample size of $N = 24$ participants ($\alpha = .05$, $1-\beta = .95$).
}
By listing this experiment on our intern study list as well as advertising through the E-Mail newsletter of the university, we were able to recruit the necessary amount of participants. 
Participants who smoked excessively ($>5$ cigarettes a day) (cf. \cite{Kirschbaum1992, Badrick2007}), were pregnant, or took cortisol-containing medication were excluded from the study.
\Edit{
In total, thirty-five healthy participants were enrolled to account for potential dropouts and to increase robustness given individual differences in cybersickness susceptibility.
}

Some participants' datasets had to be removed during data analysis due to incompleteness and other issues. 
For example, one participant was excluded because their saliva sample during the control condition was a statistically significant outlier.
This participant appeared to be exhausted and stressed, which aligns with the statistical analysis. 
The final sample comprised $N = 30$ participants (male/female: 15/15), aged 19-44 years ($M = 24.9$; $SD= 4.6$).

\Edit{
Susceptibility to visually induced motion sickness was assessed using the \ac{VIMSSQ}. The sample showed substantial interindividual variability (VIMSSQ total score: $M = 32.6$, $SD = 23.1$, range: 1.1--103.9), indicating a heterogeneous group spanning low to highly susceptible individuals rather than a sample skewed toward extreme susceptibility.
}

Each participant gave informed consent and was compensated 60\euro for participating or had the chance to acquire study credits. 
The ethics committee of Trier University approved this study. 

\section{Results\label{sec_results}}
Two-factorial repeated-measure ANOVAs were conducted to test for differences in self-reported motion sickness (\ac{SSQ}), \ac{FMS} scores, salivary cortisol, and salivary alpha-amylase in both conditions (study vs. control) and over the sampling point (time). 
For \ac{HR}, \ac{HRV}, and \ac{EDA}, one-factorial ANOVAs were calculated to test for linear and quadratic effects between baseline, motion, and recovery periods. 
Effect sizes for all ANOVAs are reported as partial $\eta^2$. 
To explain variance regarding the psychophysiological response within the study condition, separate multiple regression analyses were calculated with the predictors: age, sex, termination of motion, and \ac{SSQ}. 
All statistical analyses were done in IBM SPSS Statistics for Mac OS (Version 30.0). 
All figures were created in GraphPad Prism (version 10.4.1, San Diego, CA). 
The level of significance was set to $p < .05$. 

\subsection{Subjective Responses}
\begin{figure}
    \centering
    \includegraphics[width = .8\columnwidth]{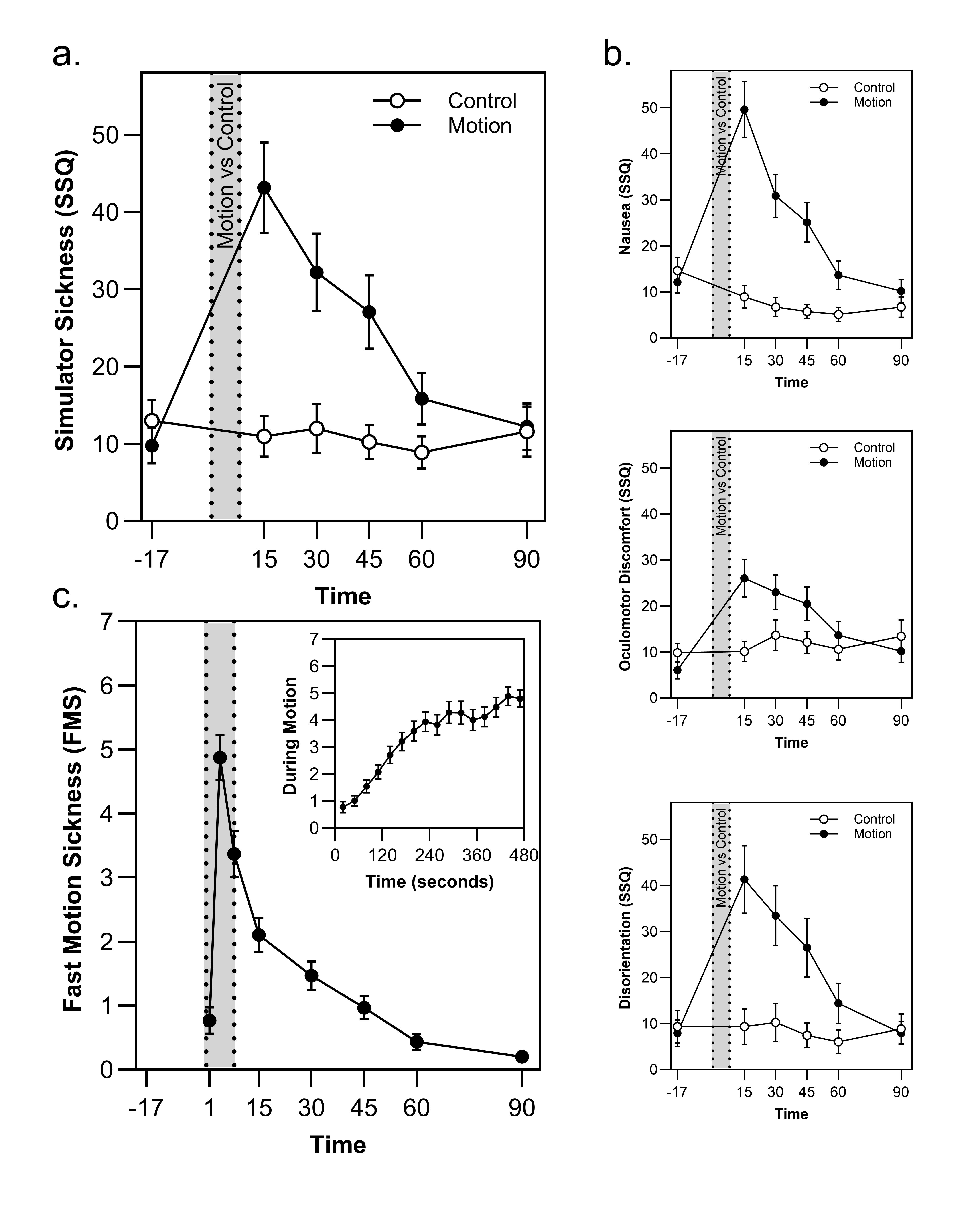}
    \Edit{
\caption{
Simulator sickness responses measured during the experiment. 
(A) Total \ac{SSQ} score assessed at each measurement time point for both the study (motion) and control sessions. 
(B) \ac{SSQ} subscales: Nausea (SSQ-N), Oculomotor Discomfort (SSQ-O), and Disorientation (SSQ-D), shown across all measurement time points for both sessions. 
(C) \ac{FMS} ratings collected every 30 seconds during the VR motion exposure only (study session). The inset shows the temporal progression of \ac{FMS} scores on a scale from 0 (no sickness) to 10 (severe sickness) during the up-to-eight-minute stimulation period. 
Bars represent mean $\pm$ SE.
\label{fig_ssq}
}}
\end{figure}

\paragraph{Simulator Sickness Questionnaire (SSQ)}
The \ac{SSQ} scores over the course of the two conditions are depicted in Fig.~\ref{fig_ssq}a.
The repeated measures ANOVA revealed a significant main effect of condition, indicating that the \ac{SSQ} significantly differed between the study and control conditions. 
Specifically, participants in the study condition exhibited significantly higher simulator sickness scores compared to those in the control condition ($F(1,29) = 17.30$, $p < .001$, $\eta^2 = .374$). 
A significant main effect of time was also observed, demonstrating that simulator sickness scores varied across the six experimental time points ($F(5,145) = 19.15$, $p < .001$,  $\eta^2 = .398$). Additionally, the interaction between condition and time was statistically significant ($F(5,145) = 22.19$, $p < .001$, $\eta^2 = .433$), suggesting that the effect of motion on simulator sickness evolved.

Differential condition effects were also shown to a similar extent for the three \ac{SSQ} subscales nausea ($F(5,145) = 28.69$, $p < .001$, $\eta^2 = .497$), oculomotor discomfort ($F(5,145) = 11.96$, $p < .001$, $\eta^2 = .292$) and disorientation ($F(5,145) = 14.18$, $p < .001$, $\eta^2 = .328$) - see Fig.~\ref{fig_ssq}b.

\paragraph{Fast Motion Sickness (FMS)}
In addition to the repeated assessments of the \ac{SSQ}, motion sickness was assessed with the \ac{FMS} (exclusively in the study condition). 
Throughout the stimulation, the subjects were asked for a rating every 30 seconds and, on average, showed the expected increase. 
At an intensity of 7 (on a scale of 0-10), the stimulation was terminated (see Section~\ref{sec_fms}). 
Following the stimulation, the values then returned to the initial level after 45 minutes at the latest, as expected ($F(6,174) = 36.90$, $p < .001$, $\eta^2 = .775$) - see Fig.~\ref{fig_ssq}c.  

These findings indicate that exposure to motion leads to a progressive increase in simulator sickness, which peaks during the exposure as indicated by the \ac{FMS} ratings and develops differently compared to the control condition as indicated by the \ac{SSQ} ratings over the whole experiment (see Fig.~\ref{fig_ssq}).

\subsection{Salivary Cortisol and Alpha-Amylase}
\begin{figure}
    \centering
    \includegraphics[width = 1.\columnwidth]{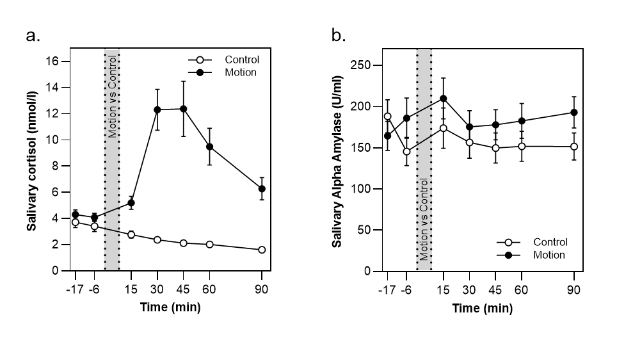}
    \caption{Salivary cortisol and alpha amylase in response to motion in VR. Bars represent mean $+/-$ SE\label{fig_cortisol}}
\end{figure}

The salivary cortisol and alpha amylase levels over the course of the two conditions are depicted in Fig.~\ref{fig_cortisol}.

\paragraph{Salivary Cortisol Levels} The repeated measures ANOVA revealed a significant main effect of condition, indicating that motion sickness had a statistically significant impact on salivary cortisol levels. 
Specifically, the analysis showed that participants exhibited significantly higher cortisol levels in the study condition compared to the control condition ($F(1,29) = 38.86$, $p < .001$, $\eta^2 = .573$). 
A significant main effect of the sample was observed, demonstrating that cortisol levels varied across different samples ($F(6,174) = 12.676$, $p < .001$, $\eta^2 = .304$). 
Additionally, the interaction between condition and sample was statistically significant ($F(6,174) = 18.622$, $p < .001$, $\eta^2 = .391$), indicating that the effect of motion sickness on cortisol levels differed depending on the sample. 
These findings suggest that exposure to motion sickness induces a pronounced \ac{HPA} axis response, as reflected in elevated cortisol levels (see Fig. \ref{fig_cortisol}a). 
In the study condition, the average response magnitude was $M=10.51 nmol/l$ ($SD=11.52 nmol/l$). 
Seventy percent of the participants showed an increase of at least $2.5 nmol/l$ and could thus be classified as responders (cf. \cite{miller2013classification, kudielka2007ten}).

\paragraph{Alpha Amylase}
The repeated measures ANOVA also revealed a significant main effect of condition for salivary alpha-amylase levels, indicating that motion sickness had a statistically significant impact ($F(1,29) = 6.671$, $p = .015$, $\eta^2 = .187$). 
Participants exhibited significantly higher alpha-amylase levels in the study condition compared to the control condition. 
A significant main effect of the sample was found ($F(6,174) = 2.733$, $p = .015$, $\eta^2 = .086$), suggesting that alpha-amylase levels varied over time. 
Furthermore, the interaction between condition and sample was statistically significant ($F(6,174) = 5.182$, $p < .001$, $\eta^2 = .152$), indicating that the effect of motion sickness on alpha-amylase levels depended on the sample. 
These results suggest that motion sickness elicits a measurable adrenergic response, as reflected in increased alpha-amylase levels (see Fig. \ref{fig_cortisol}b).

\subsection{Physiological Measures}

\begin{figure}
    \centering
    \includegraphics[width = 0.7\columnwidth]{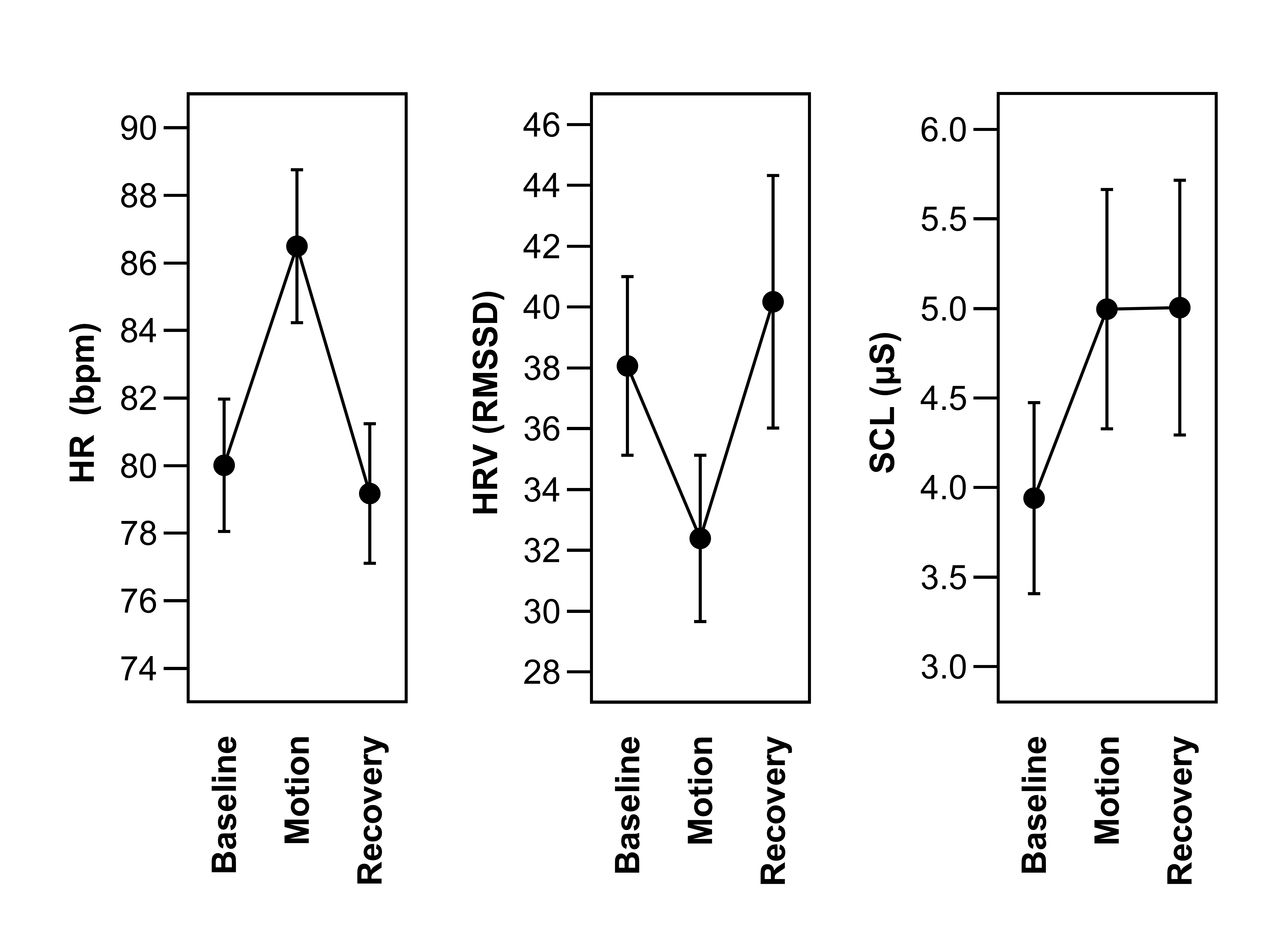}
    \caption{\acf{HR}, \acf{HRV}, and \acf{EDA} averaged for baseline, motion, and recovery phases of the study condition. Bars represent mean $+/-$ SE\label{fig_physiol}}
\end{figure}

\begin{table}
    \centering
    \tiny
    \caption{Statistical results of the ANOVAs and the planned quadratic contrasts on heart rate, heart rate variability, and skin conductance level during baseline, motion, and rest phases of the study condition.\label{tab_heartRate}}
    \begin{tabular}{lcccccccccc}
        \toprule
        & \multicolumn{2}{c}{Baseline} & \multicolumn{2}{c}{Motion} & \multicolumn{2}{c}{Recovery} & \multicolumn{2}{c}{ANOVA} & \multicolumn{2}{c}{Quadratic Contrast} \\
        \cmidrule(lr){2-3} \cmidrule(lr){4-5} \cmidrule(lr){6-7} \cmidrule(lr){8-9} \cmidrule(lr){10-11}
        & $M$ & $SD$ & $M$ & $SD$ & $M$ & $SD$ & $F$ & $p$ & $F$ & $p$ \\
        \midrule
        Heart rate (bpm) (n=28) & 80.0 & 10.4 & 86.5 & 12.0 & 79.2 & 11.0 & 30.10 & $<$.001 & 71.79 & $<$.001 \\
        HRV (RMSSD) (n=28) & 38.1 & 15.6 & 32.4 & 14.5 & 40.2 & 22.0 & 8.48 & $<$.001 & 24.19 & $<$.001 \\
        Skin conductance level (n=26) & 3.98 & 2.82 & 5.00 & 3.47 & 5.01 & 3.70 & 9.70 & $<$.001 & 14.14 & $<$.001 \\
        \bottomrule
    \end{tabular}
\end{table}

Within-subject effects of the sickness stimulus on \ac{HR}, \ac{HRV}, and \ac{EDA} were tested using separate repeated measures ANOVAs with a planned quadratic contrast to specifically test for effects during motion compared to baseline and recovery periods. Descriptives and results of the ANOVAs are found in Fig.~\ref{fig_physiol} and Table~\ref{tab_heartRate}.

In sum, the stimulation significantly increased \ac{HR}, increased \ac{EDA}, and significantly decreased \ac{HRV} (as indexed by the RMSSD), indicating that the study condition induced measurable activation of the cardiovascular system, autonomic activation, and specifically a reduction in parasympathetic tone as indicated by the effect on \ac{HRV} RMSSD.

In all three measures, the quadratic contrast suggested u-shaped effects over the experiment, however, skin conductance did not decrease during the recovery period.

\subsection{Working Memory Performance}
\begin{figure}
    \centering
    \includegraphics[width = 1.\columnwidth]{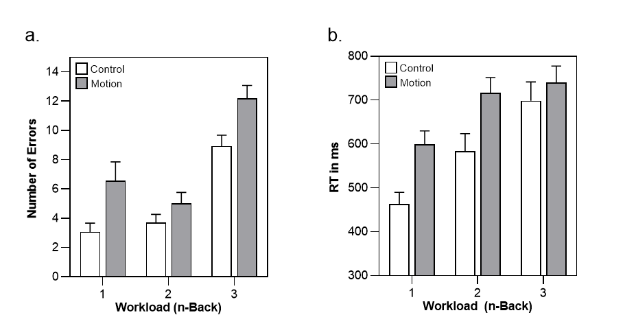}
    \caption{Errors a) and reaction time (RT) b) as a function of condition and workload in the \emph{n-Back} task. Bars represent mean $+/-$ SE\label{fig_nback}}
\end{figure}
Performance in the \emph{n-Back} task was tested based on error rates and reaction times. Descriptive values are found in Fig.~\ref{fig_nback}.

\paragraph{Errors in the n-Back task}
The ANOVA revealed a significant main effect of condition, indicating that VR-induced motion sickness had a statistically significant impact on the number of errors made. 
Specifically, the study condition compared to the control condition ($F(1,27) = 10.83$, $p = .003$, $\eta^2 = .286$). 
A highly significant main effect of cognitive load was observed, demonstrating that increasing cognitive load strongly influenced task performance ($F(2,54) = 64.37$, $p < .001$, $\eta^2 = .704$). 
Participants made more errors as the cognitive load increased, confirming that higher task difficulty impaired performance.
The interaction between condition and cognitive load was not statistically significant ($F(2,54) = 1.04$, $p = .362$, $\eta^2 = .037$).
These findings suggest that while both VR-induced motion sickness and cognitive load independently affect cognitive performance, the effects of motion sickness become more pronounced as task difficulty increases.

\paragraph{Reaction times (RT)}
 For \ac{RT}, the ANOVAs results showed a significant main effect of condition, $F(1,27) = 11.303$, $p = .002$, $\eta^2 = .295$, indicating that participants in the study condition exhibited longer RT compared to those in the control condition. 
 Additionally, there was a strong and highly significant main effect of cognitive load, $F(2,54) = 31.887$, $p < .001$, $\eta^2 = .541$, with \ac{RT}s increasing as cognitive load intensified. 
 However, the interaction between condition and load did not reach statistical significance, $F(2,54) = 2.338$, $p = .106$, $\eta^2 = .080$, suggesting that the effect of cybersickness on \ac{RT}s remained stable across different levels of cognitive load. 
 These findings indicate that cybersickness leads to prolonged reaction times, and higher cognitive load further increases \ac{RT}s, but the impact of motion sickness does not vary significantly depending on load level.

\subsection{Predictors of the Physiological Response to Motion in VR}
\begin{figure}
    \centering
    \includegraphics[width = .5\columnwidth]{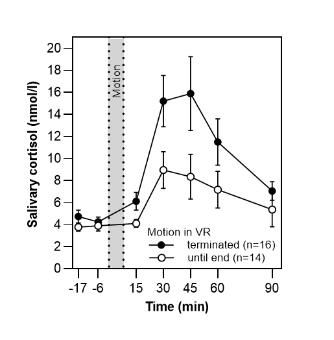}
    \caption{Salivary cortisol response to motion induction in VR for the subgroup that terminated the exposure before the end and for the subgroup with the full exposure of 8 minutes. Error bars represent SE\label{fig_subgroup}}
\end{figure}
\begin{table}
    \centering
    \caption{Linear multiple regression analyses on the physiological measures. Note. Sex was coded (1=male; 2=female) and Terminated was coded (0=no; 1=yes)\label{tab_regression}}
    \begin{minipage}{\textwidth}
    \centering
    \begin{tabular}{lccc}
        \toprule
        & $b$ & $t$ & $p$ \\
        \midrule
        \multicolumn{4}{l}{\textbf{Cortisol} ($R^2 = .243$)} \\
        Constant &  & 1.525 & .140 \\
        Sex & -0.295 & -1.694 & .103 \\
        Age & -0.022 & -0.126 & .901 \\
        Terminated & 0.404 & 2.109 & .045* \\
        SSQ & -0.299 & -1.555 & .132 \\
        \midrule
        \multicolumn{4}{l}{\textbf{a-Amylase} ($R^2 = .139$)} \\
        Constant &  & 0.037 & .971 \\
        Sex & 0.128 & 0.690 & .497 \\
        Age & 0.080 & 0.431 & .670 \\
        Terminated & 0.204 & 0.996 & .329 \\
        SSQ & -0.381 & -1.818 & .081 \\
        \midrule
        \multicolumn{4}{l}{\textbf{Heart rate} ($R^2 = .164$)} \\
        Constant &  & -0.196 & .846 \\
        Sex & 0.267 & 1.398 & .176 \\
        Age & 0.152 & 0.796 & .434 \\
        Terminated & -0.280 & -1.351 & .190 \\
        SSQ & 0.192 & 0.926 & .364 \\
        \midrule
        \multicolumn{4}{l}{\textbf{HRV} ($R^2 = .313$)} \\
        Constant &  & -0.907 & .374 \\
        Sex & 0.018 & 0.103 & .919 \\
        Age & 0.017 & 0.100 & .921 \\
        Terminated & 0.509 & 2.712 & .012* \\
        SSQ & -0.503 & -2.677 & .013* \\
        \midrule
        \multicolumn{4}{l}{\textbf{Skin conductance} ($R^2 = .139$)} \\
        Constant &  & 0.569 & .575 \\
        Sex & 0.142 & 0.715 & .482 \\
        Age & 0.061 & 0.304 & .764 \\
        Terminated & -0.294 & -1.346 & .192 \\
        SSQ & -0.085 & -0.389 & .701 \\
        \bottomrule
    \end{tabular}
    \end{minipage}
\end{table}

To explore potential linear associations between age, sex, termination of the VR exposure, \ac{SSQ} and physiological responses to sickness stimulation in VR, we conducted separate multiple regression analyses for each physiological measure, using the individual peak-baseline differences of the respective physiological measure as the criterion (see Table~\ref{tab_regression}). 
Overall, the only significant predictor in all analyses was the variable that coded the termination of motion in VR by the participant for the regression on cortisol. 
All other predictors were not significant. 
To illustrate the selective effect, we plotted the cortisol response over the experiment for those who terminated the VR exposure and those who were exposed the full 8 minutes – see Fig.~\ref{fig_subgroup}. 
The corresponding mixed ANOVA revealed a significant main effect of time ($F(6,168) = 15.52$, $p < .001$, $\eta^2 = .357$), a significant main effect of group ($F(1,28) = 4.218$, $p < .05$, $\eta^2 = .131$), and a significant interaction ($F(6,168) = 2.384$, $p < .05$, $\eta^2 = .078$), indicating pronounced cortisol responses in the participants who terminated the VR exposure before the regular end after $8 min$.

\section{Discussion\label{sec_discussion}}
\subsection{SSQ Progression and Implications}
First, we must acknowledge that the instrument used—the so-called Cybersicker—effectively induced cybersickness as intended. 
We observed a solid response compared to the baseline in both subjective measures, such as the \ac{SSQ} and \ac{FMS}, and objective measures, including cortisol, alpha-amylase, and, ultimately, the \emph{n-Back} task.
Fifteen minutes after the onset of stimulation, the average \ac{SSQ} score in the study condition reached $M=43.4$ ($SD = 31.6$), indicating a significantly elevated level (compared to $M=11.0$, $SD = 14.3$ in the control condition).
As expected, the results lie within a moderate range, also when compared to other studies (see \cite{SimnnVicente2024} for a systematic overview). 
For direct comparison, \cite{Kim2022} are cited again due to the close scientific alignment, reporting a total \ac{SSQ} score of $M=4.2$ ($SD=6.9$) pre-stimulation and $M=32.8$ ($SD=34.3$) post-stimulation; note that no control condition was included in their study.
This aligns with our design goal, which aimed to induce a level of cybersickness that still allowed participants to be systematically observed and interviewed after exposure.
It is plausible that the immediate post-exposure response was even higher. 
However, we deliberately chose to standardize the measurement time across all participants relative to the onset of the stimulus rather than its conclusion. 
Together with the necessary follow-up for biophysical measurements and VR offboarding, this resulted in a delay of the first \ac{SSQ} measurement post-stimulation by approximately 7 to 12 minutes, as previously mentioned—depending on whether and when the session was terminated.
The shortest exposure was recorded for a subject who reached an \ac{FMS} score of 7 in just under 3 minutes.
Therefore, any statements regarding the period immediately following the exposure and the first measurement remain speculative.

When examining the \ac{SSQ} subscales, the strongest response is found in the Nausea and Disorientation scales, consistent with other studies using exposure durations comparable to the one applied here \citep{Saredakis2020factors}.

\subsubsection{Delayed Symptoms}
One of the motivations for this study and the post-hoc analysis of cybersickness exposure was the recurring anecdotal reports of delayed symptoms—such as headaches—that appear in some individuals with a considerable time lag after exposure \citep{Zielasko2021sickness}. 
This phenomenon is particularly relevant when considering safety aspects of VR usage (see Section~\ref{sec_disc_nback}), but it also raises important methodological concerns. 
Delayed symptom onset could compromise the validity of within-subject study designs by introducing uncontrolled, higher-order carry-over effects \citep{zielasko2024carryOver}.

Indeed, we observed such delayed reactions in the \ac{SSQ} scores of several participants. Five out of thirty participants did not reach their peak \ac{SSQ} score immediately after exposure, but only at a later measurement point (2 at 30 min, 1 at 45 min, 1 at 60 min, and 1 at 90 min). 
A closer look at the subscales reveals that these delayed effects occur almost exclusively in the Oculomotor and Disorientation subscales—and even more prominently so. In total, 9 participants reached their peak value in one of these scales only after the first post-stimulus measurement (3 at 30 min, 3 at 45 min, 1 at 60 min, and 2 at 90 min for Oculomotor; and 3 at 30 min, 3 at 45 min, and 1 at 90 min for Disorientation). In summary, 10 out of 30 participants showed delayed symptoms, represented in at least one of these subscales.

In contrast, the Nausea subscale showed the strongest initial increase but consistently decreased afterward—monotonically in all but one case—indicating that it was not affected by delayed effects. As a result, the overall \ac{SSQ} trend may obscure lingering symptoms in the other two subscales. 
These findings provide empirical support—possibly for the first time—for the existence of significantly delayed symptoms, with some effects peaking up to 90 minutes or later post-exposure. 
This observation urgently calls for further investigation.

\Edit{
The broad distribution of VIMSSQ scores in the present sample further supports the interpretation that these delayed effects are not driven by a narrowly selected high-susceptibility subgroup, but rather reflect genuine interindividual variability in responses to the cybersickness stimulus.
}

It should also be noted that, due to the study design, we cannot rule out the possibility that the \emph{n-Back} task administered after the 30-minute \ac{SSQ} measurement may have interacted with or influenced the reported symptoms at 45, 60, and 90 minutes (cf. \cite{Sepich2022}). 
Replication of the results under controlled variation of this factor is therefore necessary.

\subsubsection{Time of Recovery}
Another notable observation is that, on average, participants returned to the \ac{SSQ} baseline level of the control condition after 90 minutes. 
This is a considerable duration, aligning with the few long-term observations reported in related studies (see Section~\ref{sec_rw_postexposure}) and raising concerns about washout periods in within-subjects experiments, which are typically much shorter.
However, it is important to note that the \emph{n-Back} task, as just mentioned, conducted at $t = 35$ minutes after stimulation onset (approximately 30 minutes post-stimulation), may have interacted with participants' SSQ score through items such as \textit{fullness of head}.
On the other hand, the fact that our stimulation protocol resulted in 70\% nausea-induced early terminations suggests that recovery times could be even longer under stronger stimulation conditions.
\Edit{
An important aspect in interpreting post-exposure recovery is the effective dose of cybersickness experienced by individual participants. Cybersickness severity is known to depend on the characteristics and duration of the stimulus as well as on individual susceptibility. Based on established categorizations of \ac{SSQ} scores \citep{Stanney1997}, the average sickness levels observed in the present study fall within the high to severe range. This provides a plausible explanation for the pronounced and prolonged recovery times observed in a subset of participants. At the same time, early termination of the stimulation by some participants resulted in a lower effective dose, which is consistent with reports of rapid recovery in other studies. Taken together, these findings support the notion that post-exposure recovery dynamics should be interpreted as dose-dependent rather than uniform across participants.
}

\subsection{Cortisol Reaction - Cybersicker as a Stressor?}
Based on findings from previous studies, we expected a stress response including self-reported stress, elevated salivary cortisol as well as autonomic activation \citep{Kim2022}. 
Overall, we found the expected pattern of increased subjective stress and physiological responses. 
However, given that cortisol is primarily considered a marker for social stress \citep{Dickerson2004acute}, the average magnitude of the baseline-to-peak response—approximately 300\% of baseline—is surprisingly high and has rarely been found in stress induction studies, neither with an in-person settings \citep{vonDawans2021}, nor in \ac{VR} setups \citep{Zimmer2019, linnig2025openTSST}. 
However, it should be taken into account that the individual responses in the present study vary significantly. 
Taking this high variability into account, we found a baseline-peak effect size of Cohen's $d = 0.72$, which is roughly comparable to previous studies with psychological stressors (such as the TSST) and previous studies inducing cybersickness. 

For instance, \citep{Kim2022} observed a cortisol level before cybersickness induction of $M = 7.75 \,\mu g/dl$ ($SD = 2.62$) and after stimulation of $M = 10.59 \,\mu g/dl$ ($SD = 4.12$), corresponding to an average increase of 37\%. 
Due to the lower variability, this corresponds to an effect size of Cohen's $d = 1.18$. In comparison, meta-analyses summarizing TSST studies reported overall pre-post effect sizes of approximately Hedges $g = 0.56$ - $0.92$ \citep{Goodman2017, Seel2025}, while studies using VR versions of the TSST observed slightly lower effect sizes around $d = 0.65$ \citep{Helminen2019}.  

\Edit{
Importantly, our findings are in line with recent work directly comparing motion sickness and psychological stress in terms of cortisol responses. Golding and colleagues reported that motion sickness induced by cross-coupled Coriolis motion produced substantially larger increases in salivary cortisol than the Trier Social Stress Test (TSST), despite comparable levels of self-reported stress and anxiety between conditions \citep{Golding2025cortisolMotion}. This dissociation suggests that motion sickness may act as a potent physiological stressor, engaging HPA-axis activity through mechanisms that are at least partly independent of purely psychological stress appraisal. The pronounced and prolonged cortisol elevations observed following VR-induced motion sickness in the present study are consistent with this interpretation and support the view that cybersickness-related stress responses may differ qualitatively from classic psychosocial stress paradigms.
}

The cortisol response to cybersickness induction opens up possibilities for using the Cybersicker as a stressor in future stress research. 
This might be especially interesting since cybersickness, stress, and psychological stress might be based on different mechanisms. 
While psychological stress presumably stimulates the \ac{HPA} axis by inducing a perceived threat to the psychological self, rotation-associated nausea in VR might rather represent a direct physiological threat to the individual's physical integrity. 
However, further investigations are necessary, as our study design does not rule out the possibility that the stress response was amplified by psychological factors, such as the presence of the experimenter triggering participants to feel more embarrassed about experiencing nausea in the presence of others.
A direct comparison of the stress response to psychological stress and cybersickness within the same sample in future studies could help to better understand the hypothesized differences or similarities between the two stressors.

Furthermore, we observed a delayed response of salivary cortisol compared to the \ac{SSQ}, with peaks occurring at $t = 30$ and $t = 45$ after the start of the simulation. 
This is expected, as cortisol is released with a temporal delay relative to the stressor and can be measured in saliva with an additional delay \citep{Goodman2017}. 
However, 90 minutes after the experiment began, cortisol levels had still not returned to control levels and remained almost three times higher than baseline. 
This extended recovery period may be due to the strong stress response (with a pronounced cortisol increase) in our study or related to other aspects of the Cybersicker like physiological stress responses in other systems, e.g., the cardiovascular system \citep{Kim2022, Allen2014}.

\subsection{n-Back Task and Implications for Safety with and after VR usage\label{sec_disc_nback}}
In the present study, we observed that regardless of the cognitive load of the working memory task, the induction of cybersickness led to a significant drop in performance compared to the control condition. 
One explanation for the reduced working memory performance after cybersickness induction relates to the well-established effects of delayed effects of stress-induced cortisol secretion on working memory \citep{Geissler2023, Geissler2025}. 
Presumably, elevated cortisol levels approximately 30 minutes after stress induction might have impaired working memory performance, as reflected in reduced accuracy on the \emph{n-Back} task compared to the control condition. This finding aligns with previous research demonstrating that acute stress and the subsequent glucocorticoid response can transiently disrupt prefrontal cortex functioning, which is critical for working memory processes \citep{Shields2016}. 
Cortisol’s delayed effects may interfere with neural activity in the dorsolateral prefrontal cortex, leading to diminished cognitive flexibility and attentional control \citep{Arnsten2009StressPrefrontal}. 

These results support the notion that the temporal dynamics of the cybersickness-induced stress response are crucial for understanding the temporal transition of potential VR-related cognitive impairments. 
In addition, our results suggest that even approximately 30 minutes after stimulation, cybersickness continues to have a significant negative impact on cognitive processes. 
This is further supported by subjective assessments using the \ac{SSQ}, indicating that participants, upon prompted self-reflection, are aware that they are still experiencing the aftereffects of cybersickness. 
If these two measures did not align, it would have far-reaching safety implications.

However, our experimental setup does not allow us to determine whether subjective discomfort and cognitive performance return to baseline simultaneously, necessitating further investigations. 
Nevertheless, we emphasize that cybersickness has a clearly measurable impact on well-defined cognitive variables even after exposure, which, to our knowledge, is the first time this has been explicitly demonstrated. 
This highlights the importance of seriously considering cybersickness and its potential aftereffects, particularly in safety-critical contexts such as operating vehicles, machinery, or other high-risk tasks in VR environments.

\Edit{
The observed impairments in cognitive performance following cybersickness induction are consistent with findings from studies employing other forms of symptom modulation. Previous work has shown that both central stimulation (e.g., visually or vestibularly driven perturbations; cf. \citep{Benelli2023cybersickness}) and peripheral stimulation can induce comparable effects on discomfort and cognitive performance. This suggests that the cognitive aftereffects observed in the present study may reflect a more general response to physiological and sensory stressors rather than a mechanism specific to cybersickness alone.
}

\subsection{Onset of Sickness \& Correlations}
Although the stimulation phase was not the primary focus of this investigation, several physiological signals were recorded for control purposes. These are briefly discussed below.

The \ac{FMS} scores increased throughout sickness stimulation, but the effect appeared to diminish toward the end of the exposure phase, suggesting a non-linear trajectory (see Fig.~\ref{fig_ssq}).
Physiological measures followed the general trend of a reaction: \ac{HR} increased during stimulation, while \ac{HRV} decreased, both returning to baseline levels during the recovery phase. 
\ac{EDA} also increased with stimulation but exhibited a more delayed reaction, remaining elevated throughout the recovery period. 
This general pattern is consistent with prior findings in real-world settings and VR research \citep{Dennison2016, Bos2005motion, Munafo2017TheVR}.

The \ac{SSQ} response was significantly predicted only by \ac{HRV} ($\beta = 0.509$, $t(28) = 2.712$, $p = .012$) and a binary variable coding early termination of the stimulation phase before reaching the maximum exposure duration of 8 minutes ($\beta = –0.503$, $t(28) = –2.677$, $p = .013$). 
No other physiological variable showed a significant relationship with acute \ac{SSQ} scores.
The \ac{VIMSSQ} was only predicting the \ac{HR} response ($r=.50$), and showed no significant associations with any other physiological variables.

Concerning endocrinological response (salivary cortisol), the only significant predictor in all regression models was the variable coding voluntary termination of VR motion by the participant. 
No other predictors—including \ac{HR}, \ac{HRV}, or \ac{EDA}—were statistically significant. 
This outcome is not necessarily unexpected and aligns with earlier findings, especially when considering the relatively small sample size ($N = 30$), which limits the detection of subtle effects such as those associated with sex differences \citep{Kelly2023gender, Howard2021}.

\subsection{Limitations}
In addition to the limitations already discussed, our study design presents several constraints that should be considered.
\Deleted{
A low-level (passive) control condition was chosen to account for the well-known circadian rhythm of the \ac{HPA} axis. 
Consequently, it cannot be ruled out that part of the observed response was influenced by unspecific contextual factors of the test situation, such as the VR environment or the presence of the experimenter. 
Future studies should incorporate active control conditions to better quantify the specific effects of \ac{HPA} activation induced by rotational movement in VR.
}
\Edit{
A low-level (passive), time-matched control condition was chosen to obtain a conservative baseline for endocrine measures that are highly sensitive to contextual arousal and environmental stimulation. However, this design cannot fully disentangle cybersickness-specific effects from other unspecific aspects of the test situation (e.g., VR context, experimenter presence). Future studies should therefore include an additional active VR control condition to separate VR-related activation from cybersickness-specific effects.
}
\Edit{
In addition, the exploratory design and moderate sample size limit the generalizability of effect size estimates, particularly given substantial interindividual differences in cybersickness susceptibility and recent recommendations to aim for higher statistical power in cybersickness research \citep{Weech2019}.
}

Furthermore, our stimulation lasted a maximum of eight minutes and was not designed to push participants to their individual limits. This implies that participants may not have experienced the full spectrum of cybersickness symptoms. 
A meta-analysis by \cite{Saredakis2020factors} categorized studies into exposure durations of 0–10 minutes, 10–20 minutes, and beyond 20 minutes. 
Their findings indicated that while oculomotor symptoms remained stable across these groups, nausea and disorientation peaked between 10 and 20 minutes. 
This suggests that our results may not be fully generalizable to longer exposure durations or more intense sickness conditions.

\section{Conclusion}
This study provides compelling evidence that VR-induced motion sickness elicits not only immediate discomfort but also sustained physiological stress responses and cognitive impairments. The consistent elevation of salivary cortisol and the delayed peak in subjective symptoms suggest that recovery from cybersickness is both protracted and complex. Our results underscore the importance of considering delayed and lingering effects in both experimental designs and applied VR use cases, particularly in high-stakes professional environments. Furthermore, the observed reduction in working memory performance following sickness induction highlights the potential safety implications when VR is used prior to cognitively demanding tasks. Future research should aim to disentangle the contributions of physiological and psychological stress components in cybersickness and develop guidelines for safe VR deployment in real-world settings.

\section*{Competing Interests}
The authors declare that they have no financial or non-financial conflicts of interest related to the subject matter or materials discussed in this manuscript. No funding was received for the conduct of this study.

\section*{Compliance with Ethical Standards}
This study involved human participants and was conducted in accordance with the ethical principles outlined in the Declaration of Helsinki. Informed consent was obtained from all participants prior to their involvement. The study protocol was reviewed and approved by the local ethics committee.


\bibliography{bib}

@article{Singer1998,
    author = {Michael J. Singer and Jennifer A. Ehrlich and Robert C. Allen},
    title ={{Virtual Environment Sickness: Adaptation to and Recovery From a Search Task}},   
    journal = {Proc. of the Human Factors and Ergonomics Society Annual Meeting},
    volume = {42},
    number = {21},
    pages = {1506-1510},
    year = {1998},
    doi = {10.1177/154193129804202109}
}

@article{Duzmanska2018,
  title={{Can Simulator Sickness Be Avoided? A Review on Temporal Aspects of Simulator Sickness}},
  author={Du{\.z}ma{\'n}ska, Natalia and Strojny, Pawe{\l} and Strojny, Agnieszka},
  journal={Frontiers in Psychology},
  volume={9},
  pages={2132},
  year={2018},
  publisher={Frontiers},
doi= {10.3389/fpsyg.2018.02132}
}

@article{Gavgani2017,
    title = {{Profiling Subjective Symptoms and Autonomic Changes Associated With Cybersickness}},
    journal = {Autonomic Neuroscience},
    volume = {203},
    pages = {41-50},
    year = {2017},
    doi = {10.1016/j.autneu.2016.12.004},
    author = {Alireza Mazloumi Gavgani and Keith V. Nesbitt and Karen L. Blackmore and Eugene Nalivaiko}
}

@article{Tanaka2004virtual,
  title={{Virtual Reality Environment Design of Managing Both Presence and Virtual Reality Sickness}},
  author={Tanaka, Nobuhisa and Takagi, Hideyuki},
  journal={Journal of Physiological Anthropology and Applied Human Science},
  volume={23},
  number={6},
  pages={313--317},
  year={2004},
  publisher={Japan Society of Physiological Anthropology},
doi={10.2114/jpa.23.313}
}

@Article{Kennedy1993,
  title={{Simulator Sickness Questionnaire: An Enhanced Method for Quantifying Simulator Sickness}},
  author={Kennedy, Robert S and Lane, Norman E and Berbaum, Kevin S and Lilienthal, Michael G},
  journal={Aviation Psychology},
  volume={3},
  number={3},
  pages={203--220},
  year={1993}
}

@inproceedings{zielasko2016hmdnav,
  title={{Evaluation of Hands-Free HMD-Based Navigation Techniques for Immersive Data Analysis}},
  author={Zielasko, Daniel and Horn, Sven and Freitag, Sebastian and Weyers, Benjamin and Kuhlen, Torsten W},
  booktitle={Proc. of IEEE 3DUI},
  pages={113--119},
  year={2016}
}

@inproceedings{zielasko2018dynamic,
  title={{Dynamic Field of View Reduction Related to Subjective Sickness Measures in an HMD-based Data Analysis Task}},
  author={Zielasko, Daniel and Mei{\ss}ner, Alexander and Freitag, Sebastian and Weyers, Benjamin and Kuhlen, Torsten W},
    booktitle={Proc. of IEEE VR Workshop on Everday Virtual Reality},
  year={2018}
}

@article{Rebenitsch2016,
    author = {Rebenitsch, Lisa and Owen, Charles},
    title = {{Review on Cybersickness in Applications and Visual Displays}},
    year = {2016},
    publisher = {Springer-Verlag},
    volume = {20},
    number = {2},
    journal = {Virtual Reality},
    pages = {101–125},
    numpages = {25}
}

@article{Stanney2003,
    author = {Kay M. Stanney and Kelly S. Hale and Isabelina Nahmens and Robert S. Kennedy},
    title ={{What to Expect from Immersive Virtual Environment Exposure: Influences of Gender, Body Mass Index, and Past Experience}},
    journal = {Human Factors},
    volume = {45},
    number = {3},
    pages = {504-520},
    year = {2003},
doi= {10.1518/hfes.45.3.504.27254}
}

@article{Stanney1997,
  title={{The Psychometrics of Cybersickness}},
  author={Stanney, Kay M and Kennedy, Robert S},
  journal={Communications of the ACM},
  volume={40},
  number={8},
  pages={66--68},
  year={1997}
}

@article{LaViola2000,
    author = {LaViola, Joseph J.},
    title = {{A Discussion of Cybersickness in Virtual Environments}},
    year = {2000},
    issue_date = {Jan. 2000},
    volume = {32},
    number = {1},
    journal = {ACM SIGCHI Bulletin},
    pages = {47–56},
    numpages = {10}
}

@article{Ilyas2012,
  title={{Simulator Sickness: A Threat to Simulator Training}},
  author={Ilyas, Rabihah},
  year={2012},
  journal={Occupational Safety and Health},
  volume = {9},
  number = {1},
}

@article{Cobb1999,
  title={{Virtual Reality-Induced Symptoms and Effects (VRISE)}},
  author={Cobb, Sue VG and Nichols, Sarah and Ramsey, Amanda and Wilson, John R},
  journal={Presence: Teleoperators \& Virtual Environments},
  volume={8},
  number={2},
  pages={169--186},
  year={1999},
  publisher={MIT Press},
doi = {10.1162/105474699566152}
}

@inproceedings{Wang2019,
  title={{User Adaptation to Cybersickness in Virtual Reality: A Qualitative Study}},
  author={Wang, Guan and Suh, Ayoung},
  year={2019},
  booktitle={Proc. of European Conference on Information Systems}
}

@article{Tyrrell2018,
  title={{Simulator Sickness in Patients With Neck Pain and Vestibular Pathology During Virtual Reality Tasks}},
  author={Tyrrell, Ryan and Sarig-Bahat, Hilla and Williams, Katrina and Williams, Grace and Treleaven, Julia},
  journal={Virtual Reality},
  volume={22},
  number={3},
  pages={211--219},
  year={2018},
  publisher={Springer},
doi={10.1007/s10055-017-0324-1}
}

@INPROCEEDINGS{Zielasko2021sickness,
  author={Zielasko, Daniel},
  booktitle={Proc. of IEEE Conference on Virtual Reality and 3D User Interfaces Abstracts and Workshops}, 
  title={{Subject 001 - A Detailed Self-Report of Virtual Reality Induced Sickness}}, 
  year={2021},
  pages={165-168},
  doi={10.1109/VRW52623.2021.00038}
}

@article{Kennedy2000duration,
  title={{Duration and Exposure to Virtual Environments: Sickness Curves During and Across Sessions}},
  author={Kennedy, Robert S and Stanney, Kay M and Dunlap, William P},
  journal={Presence: Teleoperators \& Virtual Environments},
  volume={9},
  number={5},
  pages={463--472},
  year={2000},
  publisher={MIT Press},
doi={10.1162/105474600566952}
}

@inproceedings{Serge2015simulator,
  title={{Simulator Sickness and the Oculus Rift: A First Look}},
  author={Serge, Stephen R and Moss, Jason D},
  booktitle={Proc. of Human Factors and Ergonomics Society Annual Meeting},
  pages={761--765},
  year={2015},
  organization={SAGE Publications},
doi = {10.1177/1541931215591236}
}

@article{Risi2019effects,
  title={{Effects of Postural Stability, Active Control, Exposure Duration and Repeated Exposures on Hmd Induced Cybersickness}},
  author={Risi, Dante and Palmisano, Stephen},
  journal={Displays},
  volume={60},
  pages={9--17},
  year={2019},
  publisher={Elsevier},
doi={10.1016/j.displa.2019.08.003}
}

@article{Keshavarz2011validating,
  title={{Validating an Efficient Method to Quantify Motion Sickness}},
  author={Keshavarz, Behrang and Hecht, Heiko},
  journal={Human Factors},
  volume={53},
  number={4},
  pages={415--426},
  year={2011},
  publisher={Sage Publications},
doi= {10.1177/0018720811403736}
}

@inproceedings{McHugh2019,
    author = {McHugh, Natalie and Jung, Sungchul and Hoermann, Simon and Lindeman, Robert W.},
    title = {{Investigating a Physical Dial as a Measurement Tool for Cybersickness in Virtual Reality}},
    year = {2019},
    publisher = {ACM},
    doi = {10.1145/3359996.3364259},
    booktitle = {Proc. of ACM Symposium on Virtual Reality Software and Technology},
    articleno = {32},
    numpages = {5}
}

@inproceedings{Lampton1994,
    author = {Donald R. Lampton and Eugenia M. Kolasinski and Bruce W. Knerr and James P. Bliss and John H. Bailey and Bob G. Witmer},
    title ={{Side Effects and Aftereffects of Immersion in Virtual Environments}},
    booktitle = {Proc. of the Human Factors and Ergonomics Society Annual Meeting},
    pages = {1154-1157},
    year = {1994},
    doi = {10.1177/154193129403801802},
}

@article{Saredakis2020factors,
  title={{Factors Associated With Virtual Reality Sickness in Head-Mounted Displays: A Systematic Review and Meta-Analysis}},
  author={Saredakis, Dimitrios and Szpak, Ancret and Birckhead, Brandon and Keage, Hannah AD and Rizzo, Albert and Loetscher, Tobias},
  journal={Frontiers in Human Neuroscience},
  volume={14},
  pages={96},
  year={2020},
  publisher={Frontiers Media SA},
    doi={10.3389/fnhum.2020.00096}
}

@article{Melo2018,
    title = {{Presence and Cybersickness in Immersive Content: Effects of Content Type, Exposure Time and Gender}},
    journal = {Computers \& Graphics},
    volume = {71},
    pages = {159-165},
    year = {2018},
    doi = {10.1016/j.cag.2017.11.007},
    author = {Miguel Melo and José Vasconcelos-Raposo and Maximino Bessa},
}

@Article{Thorp2022,
    AUTHOR = {Thorp, Sebastian and Sævild Ree, Alexander and Grassini, Simone},
    TITLE = {{Temporal Development of Sense of Presence and Cybersickness during an Immersive VR Experience}},
    JOURNAL = {Multimodal Technologies and Interaction},
    VOLUME = {6},
    YEAR = {2022},
    NUMBER = {5},
    ARTICLE-NUMBER = {31},
    DOI = {10.3390/mti6050031}
}

@article{Sinitski2018,
    title = {{Postural Stability and Simulator Sickness After Walking on a Treadmill in a Virtual Environment With a Curved Display}},
    journal = {Displays},
    volume = {52},
    pages = {1-7},
    year = {2018},
    doi = {10.1016/j.displa.2018.01.001},
    author = {E.H. Sinitski and A.A. Thompson and P. Godsell and J. Honey and M. Besemann},
}

@article{Teixeira2022unexpected,
  title={{Unexpected Vection Exacerbates Cybersickness During HMD-Based Virtual Reality}},
  author={Teixeira, Joel and Miellet, Sebastien and Palmisano, Stephen},
  journal={Frontiers in Virtual Reality},
  volume={3},
  pages={860919},
  year={2022},
  publisher={Frontiers},
doi={10.3389/frvir.2022.860919}
}

@article{Hakkinen2018time,
  title={{Time Course of Sickness Symptoms With Hmd Viewing of 360-Degree Videos}},
  author={H{\"a}kkinen, Jukka and Ohta, Fumiya and Kawai, Takashi},
  journal={Electronic Imaging},
  volume={31},
  pages={1--11},
  year={2018},
  publisher={Society for Imaging Science and Technology},
doi={10.2352/J.ImagingSci.Technol.2018.62.6.060403}
}

@article{Ungs1987,
    author = {Timothy J. Ungs},
    title ={{Simulator Induced Syndrome: Evidence for Long Term Simulator Aftereffects}},
    journal = {Proceedings of the Human Factors Society Annual Meeting},
    volume = {31},
    number = {5},
    pages = {505-509},
    year = {1987},
    doi = {10.1177/154193128703100505},
}

@article{Stoffregen2000,
    author = {Thomas A. Stoffregen and Lawrence J. Hettinger and Michael W. Haas and Merry M. Roe and L. James Smart},
    title ={{Postural Instability and Motion Sickness in a Fixed-Base Flight Simulator}},
    journal = {Human Factors},
    volume = {42},
    number = {3},
    pages = {458-469},
    year = {2000},
    doi = {10.1518/001872000779698097}
}

@article{Szpak2020exergaming,
  title={{Exergaming With Beat Saber: An Investigation of Virtual Reality Aftereffects}},
  author={Szpak, Ancret and Michalski, Stefan Carlo and Loetscher, Tobias},
  journal={Journal of Medical Internet Research},
  volume={22},
  number={10},
  pages={e19840},
  year={2020},
  publisher={JMIR Publications Toronto, Canada}
}

@ARTICLE{zielasko2024carryOver,
  AUTHOR={Daniel Zielasko and Ben Rehling and David Clement and Gregor Domes},   
  TITLE={{Carry-Over Effects Ruin Your (Cybersickness) Experiments and Balancing Conditions Is Not a Solution}},      
  JOURNAL={Proc. of IEEE VR Abstracts and Workshops},               
  YEAR={2024},      
    DOI={10.1109/VRW62533.2024.00007},
pages = {1--5}
}

@ARTICLE{zielasko2024cybersicker,
  AUTHOR={Daniel Zielasko and Yuen C. Law},   
  TITLE={{Cybersicker: An Open Source VR Sickness Testbed - Do you still have fun, or are you already sick}}, 
  JOURNAL={Proc. of ACM Symposium on Virtual Reality Software and Technology},               
  YEAR={2024},      
    DOI={10.1145/3641825.3689503}      
}

@article{Bubka2006Optokinetic,
  title={{Rotation Velocity Change and Motion Sickness in an Optokinetic Drum}},
  author={Bubka, Andrea and Bonato, Frederick and Urmey, Scottie and Mycewicz, Dawn},
  journal={Aviation, Space, and Environmental Medicine},
  volume={77},
  number={8},
  pages={811--815},
  year={2006},
  publisher={Aerospace Medical Association}
}

@ARTICLE{Ang2023,
    AUTHOR={Ang, Samuel  and Quarles, John},
    TITLE={{Reduction of Cybersickness in Head-Mounted Displays Use: A Systematic Review and Taxonomy of Current Strategies}},
    JOURNAL={Frontiers in Virtual Reality},
    VOLUME={4},
    YEAR={2023},
    DOI={10.3389/frvir.2023.1027552}
}

@ARTICLE{Brown2022zeroBaseline,
    AUTHOR={Brown, Phillip  and Spronck, Pieter  and Powell, Wendy }, 
    TITLE={{The Simulator Sickness Questionnaire, and the Erroneous Zero Baseline Assumption}},
    JOURNAL={Frontiers in Virtual Reality},
    VOLUME={3},
    YEAR={2022},
    DOI={10.3389/frvir.2022.945800}
}

@article{Keshavarz2023vimssq,
    author = {Behrang Keshavarz and Brandy Murovec and Niroshica Mohanathas and John F. Golding},
    title ={{The Visually Induced Motion Sickness Susceptibility Questionnaire (VIMSSQ): Estimating Individual Susceptibility to Motion Sickness-Like Symptoms When Using Visual Devices}},
    journal = {Human Factors},
    volume = {65},
    number = {1},
    pages = {107-124},
    year = {2023},
    doi = {10.1177/00187208211008687}
}

@article{Kirschbaum1992,
    title = {{‘Normal’ Cigarette Smoking Increases Free Cortisol in Habitual Smokers}},
    journal = {Life Sciences},
    volume = {50},
    number = {6},
    pages = {435-442},
    year = {1992},
    doi = {10.1016/0024-3205(92)90378-3},
    author = {C. Kirschbaum and S. Wüst and C.J. Strasburger}
}

@article{Badrick2007,
    author = {Badrick, Ellena and Kirschbaum, Clemens and Kumari, Meena},
    title = {{The Relationship Between Smoking Status and Cortisol Secretion}},
    journal = {The Journal of Clinical Endocrinology \& Metabolism},
    volume = {92},
    number = {3},
    pages = {819-824},
    year = {2007},
    month = {03},
    doi = {10.1210/jc.2006-2155},
}

@article{Kirschbaum1994,
    title = {{Salivary Cortisol in Psychoneuroendocrine Research: Recent Developments and Applications}},
    journal = {Psychoneuroendocrinology},
    volume = {19},
    number = {4},
    pages = {313-333},
    year = {1994},
    doi = {10.1016/0306-4530(94)90013-2},
    author = {Clemens Kirschbaum and Dirk H. Hellhammer}
}

@article{Rizzo2017clinical,
  title={{Is Clinical Virtual Reality Ready for Primetime?}},
  author={Rizzo, Albert and Koenig, Sebastian Thomas and others},
  journal={Neuropsychology},
  volume={31},
  number={8},
  pages={877},
  year={2017},
  publisher={American Psychological Association}
}

@book{Reason1975motion,
  title={Motion sickness},
  author={Reason, James T. and Brand, Joseph John},
  year={1975},
  publisher={Academic press}
}

@article{Dennison2016,
    title = {{Use of Physiological Signals to Predict Cybersickness}},
    journal = {Displays},
    volume = {44},
    pages = {42-52},
    year = {2016},
    doi = {10.1016/j.displa.2016.07.002},
    author = {Mark S. Dennison and A. Zachary Wisti and Michael D’Zmura}
}

@ARTICLE{Keshavarz2015,
    AUTHOR={Keshavarz, Behrang  and Riecke, Bernhard E.  and Hettinger, Lawrence J.  and Campos, Jennifer L. },
    TITLE={{Vection and Visually Induced Motion Sickness: How Are They Related?}},
    JOURNAL={Frontiers in Psychology},
    VOLUME={6},
    YEAR={2015},
    DOI={10.3389/fpsyg.2015.00472}
}

@article{Parsons2018VRChallenges,
  title     = {{Virtual Reality for Research in Social Neuroscience}},
  author    = {Parsons, Thomas D. and Gaggioli, Andrea and Riva, Giuseppe},
    JOURNAL = {Brain Sciences},
    VOLUME = {7},
    YEAR = {2017},
    NUMBER = {4},
    ARTICLE-NUMBER = {42},
    DOI = {10.3390/brainsci7040042}
}

@article{Bos2005motion,
  title={{Motion Sickness Symptoms in a Ship Motion Simulator: Effects of Inside, Outside, and No View}},
  author={Bos, Jelte E and MacKinnon, Scott N and Patterson, Anthony},
  journal={{Aviation, Space, and Environmental Medicine}},
  volume={76},
  number={12},
  pages={1111--1118},
  year={2005},
  publisher={Aerospace Medical Association}
}

@article{Munafo2017TheVR,
  title={{The Virtual Reality Head-Mounted Display Oculus Rift Induces Motion Sickness and Is Sexist in Its Effects}},
  author={Justin Munafo and Meg Diedrick and Thomas A. Stoffregen},
  journal={Experimental Brain Research},
  year={2017},
  volume={235},
  pages={889-901},
}

@Article{Kim2022,
    author="Kim, Yoon Sang
    and Won, JuHye
    and Jang, Seong-Wook
    and Ko, Junho",
    title={{Effects of Cybersickness Caused by Head-Mounted Display--Based Virtual Reality on Physiological Responses: Cross-sectional Study}},
    journal="JMIR Serious Games",
    year="2022",
    month="Oct",
    day="17",
    volume="10",
    number="4",
    doi="10.2196/37938",
}

@article{McEwen2007,
    author = {McEwen, Bruce S.},
    title = {{Physiology and Neurobiology of Stress and Adaptation: Central Role of the Brain}},
    journal = {Physiological Reviews},
    volume = {87},
    number = {3},
    pages = {873-904},
    year = {2007},
    doi = {10.1152/physrev.00041.2006}
}

@article{Sandi2013,
    author = {Sandi, Carmen},
    title = {{Stress and Cognition}},
    journal = {WIREs Cognitive Science},
    volume = {4},
    number = {3},
    pages = {245-261},
    doi = {10.1002/wcs.1222},
    year = {2013}
}

@article{Schoofs2008,
    title = {{Psychosocial Stress Induces Working Memory Impairments in an n-Back Paradigm}},
    journal = {Psychoneuroendocrinology},
    volume = {33},
    number = {5},
    pages = {643-653},
    year = {2008},
    doi = {10.1016/j.psyneuen.2008.02.004},
    author = {Daniela Schoofs and Diana Preuß and Oliver T. Wolf}
}

@article{Shields2016,
    title = {{The Effects of Acute Stress on Core Executive Functions: A Meta-Analysis and Comparison With Cortisol}},
    journal = {Neuroscience \& Biobehavioral Reviews},
    volume = {68},
    pages = {651-668},
    year = {2016},
    doi = {10.1016/j.neubiorev.2016.06.038},
    author = {Grant S. Shields and Matthew A. Sazma and Andrew P. Yonelinas}
}

@article{Arnsten2009StressPrefrontal,
  title     = {{Stress Signalling Pathways That Impair Prefrontal Cortex Structure and Function}},
  author    = {Arnsten, Amy F. T.},
  journal   = {Nature Reviews Neuroscience},
  volume    = {10},
  pages     = {410--422},
  year      = {2009},
  doi       = {10.1038/nrn2648},
}

@article{Faul2007gPower,
  title={{G* Power 3: A Flexible Statistical Power Analysis Program for the Social, Behavioral, and Biomedical Sciences}},
  author={Faul, Franz and Erdfelder, Edgar and Lang, Albert-Georg and Buchner, Axel},
  journal={Behavior Research Methods},
  volume={39},
  number={2},
  pages={175--191},
  year={2007},
  publisher={Springer},
  doi={10.3758/BF03193146}
}

@article{Dressendorfer1992CortisolBiotin,
  title     = {{Synthesis of a Cortisol-Biotin Conjugate and Evaluation as a Tracer in an Immunoassay for Salivary Cortisol Measurement}},
  author    = {Dressend\"orfer, R. A. and Kirschbaum, C. and Rohde, W. and Stahl, F. and Strasburger, C. J.},
  journal   = {Journal of Steroid Biochemistry and Molecular Biology},
  volume    = {43},
  number    = {7},
  pages     = {683--692},
  year      = {1992},
  doi       = {10.1016/0960-0760(92)90294-S},
}

@article{LorentzGutschowRenner1999,
    title = {{Evaluation of a Direct $\alpha$-Amylase Assay Using 2-Chloro-4-nitrophenyl-$\alpha$-D-maltotrioside}},
    author = {Klaus Lorentz and Barbara G\"utschow and Florian Renner},
    pages = {1053--1062},
    volume = {37},
    number = {11-12},
    journal = {Clinical Chemistry and Laboratory Medicine},
    doi = {10.1515/CCLM.1999.154},
    year = {1999},
}

@article{WinnDeen1988,
    author = {Winn-Deen, E S and David, H and Sigler, G and Chavez, R},
    title = {{Development of a Direct Assay for Alpha-Amylase}},
    journal = {Clinical Chemistry},
    volume = {34},
    number = {10},
    pages = {2005-2008},
    year = {1988},
    month = {10},
    doi = {10.1093/clinchem/34.10.2005},
}

@article{Geissler2023,
    title = {{Time-Dependent Effects of Acute Stress on Working Memory Performance: A Systematic Review and Hypothesis}},
    journal = {Psychoneuroendocrinology},
    volume = {148},
    pages = {105998},
    year = {2023},
    doi = {10.1016/j.psyneuen.2022.105998},
    author = {Christoph F. Geißler and Maximilian A. Friehs and Christian Frings and Gregor Domes}
}

@article{Geissler2025,
    author = {Christoph Felix Geissler and Christian Frings and Gregor Domes},
    title = {{The Effects of Stress on Working-Memory-Related Prefrontal Processing: An fNIRS Study}},
    journal = {Stress},
    volume = {28},
    number = {1},
    pages = {2472067},
    year = {2025},
    publisher = {Taylor \& Francis},
    doi = {10.1080/10253890.2025.2472067}
}

@article{Dickerson2004acute,
  title={{Acute Stressors and Cortisol Responses: A Theoretical Integration and Synthesis of Laboratory Research}},
  author={Dickerson, Sally S and Kemeny, Margaret E},
  journal={Psychological bulletin},
  volume={130},
  number={3},
  pages={355},
  year={2004},
  publisher={American Psychological Association},
    doi={10.1037/0033-2909.130.3.355}
}

@article{ulrich2009neural,
  title={{Neural Regulation of Endocrine and Autonomic Stress Responses}},
  author={Ulrich-Lai, Yvonne M and Herman, James P},
  journal={Nature Reviews Neuroscience},
  volume={10},
  number={6},
  pages={397--409},
  year={2009},
  publisher={Nature Publishing Group UK London},
doi={10.1038/nrn2647}
}

@article{miller2013classification,
  title={{Classification Criteria for Distinguishing Cortisol Responders From Nonresponders to Psychosocial Stress: Evaluation of Salivary Cortisol Pulse Detection in Panel Designs}},
  author={Miller, Robert and Plessow, Franziska and Kirschbaum, Clemens and Stalder, Tobias},
  journal={Psychosomatic Medicine},
  volume={75},
  number={9},
  pages={832--840},
  year={2013},
  publisher={LWW},
doi={10.1097/PSY.0000000000000002}
}

@article{kudielka2007ten,
  title={{Ten years of research with the Trier Social Stress Test (TSST)-revisited}},
  author={Kudielka, Brigitte M and Hellhammer, Dirk H and Kirschbaum, Clemens},
  journal={Social neuroscience: Integrating biological and psychological explanations of social behavior},
  pages={56--83},
  year={2007},
  publisher={Guilford Press New York}
}

@article{vonDawans2021,
    title = {{The Effects of Acute Stress and Stress Hormones on Social Cognition and Behavior: Current State of Research and Future Directions}},
    journal = {Neuroscience \& Biobehavioral Reviews},
    volume = {121},
    pages = {75-88},
    year = {2021},
    doi = {10.1016/j.neubiorev.2020.11.026},
    author = {Bernadette {von Dawans} and Julia Strojny and Gregor Domes}
}

@article{Zimmer2019,
    title = {{Virtually Stressed? A Refined Virtual Reality Adaptation of the Trier Social Stress Test (Tsst) Induces Robust Endocrine Responses}},
    journal = {Psychoneuroendocrinology},
    volume = {101},
    pages = {186-192},
    year = {2019},
    doi = {10.1016/j.psyneuen.2018.11.010},
    author = {Patrick Zimmer and Benjamin Buttlar and Georg Halbeisen and Eva Walther and Gregor Domes}
}

@article{linnig2025openTSST,
  title     = {{Open TSST VR: Psychobiological reactions to an open version of the Trier Social Stress Test in Virtual Reality}},
  author    = {Katrin Linnig and Saskia Seel and Bernadette von Dawans and William Standard and Daniel Zielasko and Benjamin Weyers and Gregor Domes},
  journal   = {Behavior Research Methods},
  year      = {2025},
    doi={10.3758/s13428-025-02662-x}
}

@article{Goodman2017,
    title = {{Meta-Analytical Assessment of the Effects of Protocol Variations on Cortisol Responses to the Trier Social Stress Test}},
    journal = {Psychoneuroendocrinology},
    volume = {80},
    pages = {26-35},
    year = {2017},
    doi = {10.1016/j.psyneuen.2017.02.030},
    author = {William K. Goodman and Johanna Janson and Jutta M. Wolf}
}

@article{Seel2025,
    title = {{Experimental Stress Induction in Children and Adolescents With the Trier Social Stress Test (Tsst): A Systematic Review and Meta-Analysis}},
    journal = {Psychoneuroendocrinology},
    volume = {177},
    pages = {107454},
    year = {2025},
    doi = {10.1016/j.psyneuen.2025.107454},
    author = {Saskia Seel and Bernhard Pastötter and Gregor Domes}
}

@article{Helminen2019,
    title = {A meta-analysis of cortisol reactivity to the Trier Social Stress Test in virtual environments},
    journal = {Psychoneuroendocrinology},
    volume = {110},
    pages = {104437},
    year = {2019},
    doi = {10.1016/j.psyneuen.2019.104437},
    author = {Emily C. Helminen and Melissa L. Morton and Qiu Wang and Joshua C. Felver}
}

@article{Allen2014,
    title = {{Biological and Psychological Markers of Stress in Humans: Focus On the Trier Social Stress Test}},
    journal = {Neuroscience \& Biobehavioral Reviews},
    volume = {38},
    pages = {94-124},
    year = {2014},
    doi = {10.1016/j.neubiorev.2013.11.005},
    author = {Andrew P. Allen and Paul J. Kennedy and John F. Cryan and Timothy G. Dinan and Gerard Clarke}
}

@article{Recenti2021toward,
  title={{Toward Predicting Motion Sickness Using Virtual Reality and a Moving Platform Assessing Brain, Muscles, and Heart Signals}},
  author={Recenti, Marco and Ricciardi, Carlo and Aubonnet, Romain and Picone, Ilaria and Jacob, Deborah and Svansson, Halld{\'o}r {\'A}R and Agnarsd{\'o}ttir, S{\'o}lveig and Karlsson, Gunnar H and Baeringsd{\'o}ttir, Vald{\'\i}s and Petersen, Hannes and others},
  journal={Frontiers in Bioengineering and Biotechnology},
  volume={9},
  pages={635661},
  year={2021},
  publisher={Frontiers Media SA},
doi = {10.3389/fbioe.2021.635661}
}

@inproceedings{Lin2018,
  title={{65-3: The Quantization of Cybersickness Level Using EEG and ECG for Virtual Reality Head-Mounted Display}},
  author={Lin, Yi-Tien and Chien, Yu-Yi and Wang, Hsiao-Han and Lin, Fang-Cheng and Huang, Yi-Pai},
  booktitle={SID Symposium Digest of Technical Papers},
  volume={49},
  number={1},
  pages={862--865},
  year={2018},
  organization={Wiley Online Library},
doi={10.1002/sdtp.12267}
}

@INPROCEEDINGS{Islam2020,
  author={Islam, Rifatul and Lee, Yonggun and Jaloli, Mehrad and Muhammad, Imtiaz and Zhu, Dakai and Rad, Paul and Huang, Yufei and Quarles, John},
  booktitle={IEEE International Symposium on Mixed and Augmented Reality}, 
  title={{Automatic Detection and Prediction of Cybersickness Severity using Deep Neural Networks from user’s Physiological Signals}}, 
  year={2020},
  pages={400-411},
  doi={10.1109/ISMAR50242.2020.00066}
}

@article{Li2020,
    title = {{VR Motion Sickness Recognition by Using EEG Rhythm Energy Ratio Based on Wavelet Packet Transform}},
    journal = {Computer Methods and Programs in Biomedicine},
    volume = {188},
    pages = {105266},
    year = {2020},
    doi = {10.1016/j.cmpb.2019.105266},
    author = {Xiaolu Li and Changrong Zhu and Cangsu Xu and Junjiang Zhu and Yuntang Li and Shanqiang Wu}
}

@ARTICLE{Liao2020,
  author={Liao, Chung-Yen and Tai, Shao-Kuo and Chen, Rung-Ching and Hendry, Hendry},
  journal={IEEE Access}, 
  title={{Using EEG and Deep Learning to Predict Motion Sickness Under Wearing a Virtual Reality Device}}, 
  year={2020},
  volume={8},
  pages={126784-126796},
  doi={10.1109/ACCESS.2020.3008165}
}

@InProceedings{Kim2019,
    author = {Kim, Jinwoo and Kim, Woojae and Oh, Heeseok and Lee, Seongmin and Lee, Sanghoon},
    title = {{A Deep Cybersickness Predictor Based on Brain Signal Analysis for Virtual Reality Contents}},
    booktitle = {Proceedings of the IEEE\/CVF International Conference on Computer Vision (ICCV)},
    year = {2019},
doi={10.1109/ICCV.2019.01068}
}

@INPROCEEDINGS{Magaki2019,
  author={Magaki, Takurou and Vallance, Michael},
  booktitle={IEEE Conference on Virtual Reality and 3D User Interfaces (VR)}, 
  title={Developing an Accessible Evaluation Method of VR Cybersickness}, 
  year={2019},
  pages={1072-1073},
  doi={10.1109/VR.2019.8797748}
}

@article{Garcia2019development,
  title={{Development of a Classifier to Determine Factors Causing Cybersickness in Virtual Reality Environments}},
  author={Garcia-Agundez, Augusto and Reuter, Christian and Becker, Hagen and Konrad, Robert and Caserman, Polona and Miede, Andr{\'e} and G{\"o}bel, Stefan},
  journal={Games for Health Journal},
  volume={8},
  number={6},
  pages={439--444},
  year={2019},
  publisher={Mary Ann Liebert, Inc.},
doi={10.1089/g4h.2019.0045}
}

@article{oh2021machine,
  title={{Machine--Deep--Ensemble Learning Model for Classifying Cybersickness Caused by Virtual Reality Immersion}},
  author={Oh, SeungJun and Kim, Dong-Keun},
  journal={Cyberpsychology, Behavior, and Social Networking},
  volume={24},
  number={11},
  pages={729--736},
  year={2021},
  publisher={Mary Ann Liebert, Inc.},
doi = {10.1089/cyber.2020.0613}
}

@inproceedings{Dennison2019,
    author = {Mark S. Dennison and Mike D'Zmura and Andre Harrison and Michael Lee and Adrienne Raglin},
    title = {{Improving Motion Sickness Severity Classification Through Multi-Modal Data Fusion}},
    volume = {11006},
    booktitle = {Artificial Intelligence and Machine Learning for Multi-Domain Operations Applications},
    organization = {International Society for Optics and Photonics},
    publisher = {SPIE},
    pages = {110060T},
    year = {2019},
    doi = {10.1117/12.2519085}
}

@INPROCEEDINGS{Islam2022towads,
  author={Islam, Rifatul and Desai, Kevin and Quarles, John},
  booktitle={IEEE International Symposium on Mixed and Augmented Reality (ISMAR)}, 
  title={{Towards Forecasting the Onset of Cybersickness by Fusing Physiological, Head-tracking and Eye-tracking with Multimodal Deep Fusion Network}}, 
  year={2022},
  pages={121-130},
  doi={10.1109/ISMAR55827.2022.00026}
}

@ARTICLE{Adhikari2022,
  author={Ashu Adhikari and Daniel Zielasko and Ivan Aguilar and Alexander Bretin and Ernst Kruijff and Markus von der Heyde and Bernhard E. Riecke},
  journal={IEEE Transactions on Visualization and Computer Graphics}, 
  title={{Integrating Continuous and Teleporting VR Locomotion into a Seamless “HyperJump” Paradigm}}, 
  year={2022},
  doi={10.1109/TVCG.2022.3207157}
}

@INPROCEEDINGS{Kelly2023gender,
  author={Kelly, Jonathan W. and Gilbert, Stephen B. and Dorneich, Michael C. and Costabile, Kristi A.},
  booktitle={2023 IEEE Conference on Virtual Reality and 3D User Interfaces Abstracts and Workshops (VRW)}, 
  title={{Gender Differences in Cybersickness: Clarifying Confusion and Identifying Paths Forward}}, 
  year={2023},
  pages={283-288},
  doi={10.1109/VRW58643.2023.00067}
}

@article{Howard2021,
    author = {Howard, Matt C. and Van Zandt, Elise C.},
    title = {{A Meta-Analysis of the Virtual Reality Problem: Unequal Effects of Virtual Reality Sickness Across Individual Differences}},
    year = {2021},
    publisher = {Springer-Verlag},
    volume = {25},
    number = {4},
    doi = {10.1007/s10055-021-00524-3},
    journal = {Virtual Reality},
    pages = {1221–1246},
    numpages = {26},
}

@article{SimnnVicente2024,
    title = {{Cybersickness. a Systematic Literature Review of Adverse Effects Related to Virtual Reality}},
    journal = {Neurología},
    volume = {39},
    number = {8},
    pages = {701-709},
    year = {2024},
    doi = {https://doi.org/10.1016/j.nrl.2022.04.009},
    author = {L. Simón-Vicente and S. Rodríguez-Cano and V. Delgado-Benito and V. Ausín-Villaverde and E. {Cubo Delgado}}
}

@ARTICLE{Sepich2022, 
    AUTHOR={Sepich, Nathan C.  and Jasper, Angelica  and Fieffer, Stephen  and Gilbert, Stephen B.  and Dorneich, Michael C.  and Kelly, Jonathan W. },       
    TITLE={The impact of task workload on cybersickness},       
    JOURNAL={Frontiers in Virtual Reality},     
    VOLUME={3},
    YEAR={2022},
    DOI={10.3389/frvir.2022.943409}
}

@article{Keshavarz2022,
    title = {{Detecting and Predicting Visually Induced Motion Sickness With Physiological Measures in Combination With Machine Learning Techniques}},
    journal = {International Journal of Psychophysiology},
    volume = {176},
    pages = {14-26},
    year = {2022},
    doi = {10.1016/j.ijpsycho.2022.03.006},
    author = {Behrang Keshavarz and Katlyn Peck and Sia Rezaei and Babak Taati}
}

@ARTICLE{Weech2019,
    
    AUTHOR={Weech, Séamas  and Kenny, Sophie  and Barnett-Cowan, Michael },
    TITLE={{Presence and Cybersickness in Virtual Reality Are Negatively Related: A Review}},         
    JOURNAL={Frontiers in Psychology},          
    VOLUME={Volume 10 - 2019},
    YEAR={2019},
    DOI={10.3389/fpsyg.2019.00158}
}

@article{Benelli2023cybersickness,
  title = {{Frequency-Dependent Reduction of Cybersickness in Virtual Reality by Transcranial Oscillatory Stimulation of the Vestibular Cortex}},
  author = {Benelli, Alberto and Neri, Francesco and Cinti, Alessandra and Pasqualetti, Patrizio and Romanella, Sara M. and Giannotta, Alessandro and De Monte, David and Mandal\`{a}, Marco and Smeralda, Carmelo and Prattichizzo, Domenico and Santarnecchi, Emiliano and Rossi, Simone},
  journal = {Neurotherapeutics},
  year = {2023},
  volume = {20},
  number = {6},
  pages = {1796--1807},
  doi = {10.1007/s13311-023-01437-6}
}

@inproceedings{Golding2025cortisolMotion,
  author    = {Golding, John F. and O’Brien, Lisa-Marie and Nabongo, Mariam and Flynn, Maria and Clow, Angela and Smyth, Nina},
  title     = {{The Effects of Motion Sickness Versus Psychological Stress on Salivary Cortisol Responses}},
  booktitle = {British Society of Neuro-Otology (BSNO) Meeting Conference Abstract},
  address   = {St Thomas' Hospital, London, UK},
  year      = {2025}
}

\end{document}